\documentclass[aps,prd,twocolumn,superscriptaddress,showpacs,preprintnumbers,nofootinbib,altaffilletter,bibnotes]{revtex4}

\usepackage{amsmath}
\usepackage{amssymb}
\usepackage{epsfig}

\begin{document}
\def\simle{\mathrel{\rlap{\raise 0.511ex \hbox{$<$}}{\lower 0.511ex \hbox{$\sim$}}}}




\title{Performance Of A Liquid Argon Time Projection Chamber Exposed
To The WANF Neutrino Beam}

\newcommand{\aquila}{ Dipartimento d Fisica e INFN, Universita' dell'Aquila, via Vetoio, L'Aquila, Italy}
\newcommand{\pavia}{ Dipartimento di Fisica e INFN, Universita' di Pavia, via Bassi 6, Pavia, Italy}
\newcommand{\milano}{ Dipartimento di Fisica e INFN, Universita' di Milano, via Celoria 16, Milano, Italy}
\newcommand{\granada}{ Departamento de F\'\i sica Te\'orica y del Cosmos and
                Centro Andaluz de F\'\i sica de Part\'\i culas Elementales (CAFPE),
                 Universidad de Granada, Granada, Spain}
\newcommand{\frascati}{ Laboratori Nazionali dell'INFN di Frascati, via Fermi 40, Frascati (RM), Italy}
\newcommand{\cern}{ CERN, CH-1211 Geneva 23, Switzerland}
\newcommand{\padova}{ Dipartimento di Fisica e INFN, Universita' di Padova, via Marzolo 8, Padova, Italy}
\newcommand{\ucla}{ Department of Physics, UCLA, Los Angeles, CA 90024, USA}
\newcommand{\KATOWICE}{ Institute of Physics, University of Silesia,
                   Katowice, Poland}
\newcommand{\torin}{ ICGF-CNR, Corso Fiume 4, Torino, Italy}
\newcommand{\LNGS}{ Laboratori Nazionali del Gran Sasso (LNGS) INFN,
               Assergi, Italy}
\newcommand{\torino}{ Dipartimento di Fisica e INFN, Universita' di Torino, via Giuria 1, Torino, Italy}
\newcommand{\WARSZAWA}{ A.So\l tan Institute for Nuclear Studies, Warszawa, Poland}
\newcommand{\ethz}{ Institute for Particle Physics, ETH H\"onggerberg, Z\"urich, Switzerland}
\newcommand{\pisa}{ INFN Pisa, Largo B. Pontecorvo 3, Pisa, Italy}
\newcommand{\wroclaw}{ Institute of Theoretical Physics, Wroc\l aw University, Wroc\l aw, Poland}
\newcommand{\KRAKOW}{ H. Niewodnicza\'nski Institute of Nuclear Physics, Krak\'ow, Poland}

\newcommand{\yale}{Present address: Yale University.} 
\newcommand{\oakridge}{Present address: Oakridge National Laboratory.}
\newcommand{\waseda}{Present address: Waseda University.}

\affiliation{\aquila}
\affiliation{\pavia}
\affiliation{\milano}
\affiliation{\granada}
\affiliation{\frascati}
\affiliation{\cern}
\affiliation{\padova}
\affiliation{\ucla}
\affiliation{\KATOWICE}
\affiliation{\torin}
\affiliation{\LNGS}
\affiliation{\torino}
\affiliation{\WARSZAWA}
\affiliation{\ethz}
\affiliation{\pisa}
\affiliation{\wroclaw}
\affiliation{\KRAKOW}


\author{F.~Arneodo} 
\affiliation{\aquila}

\author{P.~Benetti}
\affiliation{\pavia}

\author{M.~Bonesini} 
\affiliation{\milano}

\author{A.~Borio di Tigliole}
\affiliation{\pavia}

\author{B.~Boschetti}
\affiliation{\milano}

\author{A.~Bueno}
\affiliation{\granada}

\author{E.~Calligarich} 
\affiliation{\pavia}

\author{F.~Casagrande} 
\altaffiliation{\oakridge}
\affiliation{\frascati}

\author{D.~Cavalli}
\affiliation{\milano}

\author{F.~Cavanna}
\affiliation{\aquila}

\author{P.~Cennini}
\affiliation{\cern}

\author{S.~Centro} 
\affiliation{\padova}

\author{E.~Cesana}
\affiliation{\pavia}

\author{D.~Cline} 
\affiliation{\ucla}


\author{A.~Curioni}
\altaffiliation{\yale}
\affiliation{\milano}

\author{I.~De~Mitri} 
\affiliation{\aquila}

\author{C.~De~Vecchi} 
\affiliation{\padova}
\affiliation{\pavia}

\author{R.~Dolfini} 
\affiliation{\pavia}

\author{A.~Ferrari}
\affiliation{\cern}

\author{A.~Ghezzi}
\affiliation{\milano}

\author{A .~Guglielmi} 
\affiliation{\padova}

\author{J.~Kisiel}
\affiliation{\KATOWICE}

\author{G.~Mannocchi} 
\affiliation{\torin}

\author{A.~Mart\'\i nez~de~la~Ossa}
\affiliation{\granada}

\author{C.~Matthey}
\affiliation{\ucla}

\author{F.~Mauri}
\affiliation{\pavia}

\author{C.~Montanari} 
\affiliation{\pavia}

\author{S.~Navas}
\affiliation{\granada}

\author{P.~Negri}
\affiliation{\milano}

\author{M.~Nicoletto}
\affiliation{\padova}

\author{S.~Otwinowski} 
\affiliation{\ucla}

\author{M.~Paganoni} 
\affiliation{\milano}

\author{O.~Palamara}
\affiliation{\LNGS}

\author{A.~Pepato}
\affiliation{\padova}

\author{L.~Periale}
\affiliation{\torin}

\author{G.~Piano~Mortari} 
\affiliation{\aquila}

\author{P.~Picchi}
\affiliation{\torino}
\affiliation{\frascati}

\author{F.~Pietropaolo}
\affiliation{\padova}

\author{A.~Puccini} 
\affiliation{\cern}

\author{A.~Pullia}
\affiliation{\milano}

\author{S.~Ragazzi}
\affiliation{\milano}

\author{T.~Rancati}
\affiliation{\milano}

\author{A.~Rappoldi}
\affiliation{\pavia}

\author{G.L.~Raselli} 
\affiliation{\pavia}

\author{N.~Redaelli} 
\affiliation{\milano}

\author{E.~Rondio}
\affiliation{\WARSZAWA}

\author{A.~Rubbia }
\affiliation{\ethz}

\author{C.~Rubbia} 
\affiliation{\pavia}

\author{P.R.~Sala}
\affiliation{\milano}

\author{F.~Sergiampietri} 
\affiliation{\pisa}

\author{J.~Sobczyk}
\affiliation{\wroclaw}

\author{S.~Suzuki}
\altaffiliation{\waseda}
\affiliation{\torino}
\affiliation{\frascati}

\author{T.~Tabarelli~de~Fatis}
\affiliation{\milano}

\author{M.~Terrani}
\affiliation{\pavia}

\author{F.~Terranova} 
\affiliation{\milano}

\author{A.~Tonazzo}
\affiliation{\milano}

\author{S.~Ventura}
\affiliation{\padova}

\author{C.~Vignoli}
\affiliation{\pavia}

\author{H.~Wang} 
\affiliation{\ucla}

\author{A.~Zalewska}
\affiliation{\KRAKOW}

\collaboration{The ICARUS-Milano Collaboration}
\noaffiliation


\begin{abstract}
We present the results of the first exposure of a Liquid Argon TPC to a
multi-GeV neutrino beam. The data have been collected with a 50 liters
ICARUS-like chamber located between the CHORUS and NOMAD experiments
at the CERN West Area Neutrino Facility (WANF). We discuss both the
instrumental performance of the detector and its capability to
identify and reconstruct low multiplicity neutrino interactions.
\end{abstract}


\maketitle

\section{Introduction}
\label{introduction}

The Liquid Argon Time Projection Chamber (LAr TPC) technology was
originally proposed by C. Rubbia in 1977~\cite{ref:Rubbia77}, 
as a novel concept for a massive neutrino detector that allows
three-dimensional event reconstruction with high accuracy and precise
calorimetric measurement. The feasibility of the technology, developed
inside the context of the ICARUS project~\cite{ref:proposal}, has
been demonstrated by an extensive R$\&$D programme 
that addressed the main technological challenges of the LAr
detection technique (i.e. argon purification, wire chambers, electronics, 
DAQ, etc.). The ICARUS collaboration has been able to operate prototypes of
increasing mass (3 tons~\cite{ref:3ton,ref:purification}, 14 tons~\cite{ref:14ton}, 
600 tons~\cite{ref:600ton}) and therefore the LAr TPC 
can now be considered as a mature technology
able to produce multi-kton detectors for astroparticle and neutrino
physics. 

Within the broad physics programme of the ICARUS project, an important
issue is the study of oscillations through the detection of
long-baseline neutrino interactions. The possibility to observe
$\nu_\mu \to \nu_e$ and $\nu_\mu \to \nu_\tau$ by means of kinematic
criteria is known to be limited, among others, by the knowledge we have of 
nuclear effects (Fermi motion, nuclear re-scattering and absorption, etc.). 
Therefore it is very important to acquire experimental data in order 
to tune the existing Monte-Carlo models. In addition, the exposure of a LAr TPC
to a neutrino beam is mandatory to demonstrate the high recognition 
capability of the technique and gain experience with real neutrino events. 

In 1997, the ICARUS collaboration together with a group from 
INFN and Milano University~\cite{ref:Stefano} proposed to expose a 50 liter LAr TPC 
to the multi-GeV wide band neutrino beam of the CERN West Area Neutrino Facility
(WANF)~\cite{ref:wanf}, during the NOMAD~\cite{ref:NOMAD} and
CHORUS~\cite{ref:CHORUS} data taking. The test was part of an 
R\&D program for a medium baseline
$\nu_\tau$ appearance experiment~\cite{ref:icarus-cern-mi}. The idea
was to collect a substantial sample of quasi-elastic interactions 
($\nu_\mu$ + n $\to$ p + $\mu^-$) to study the following physics items: 
\begin{itemize} 
\item Measurement of the acoplanarity and missing transverse momentum
in events with the $\mu$-p topology in the final state, in order to
assess Fermi motion and proton re-scattering inside the nucleus. 
\item Appearance of nuclear fragments (short tracks and {\it blobs}
around the primary interaction vertex) in quasi-elastic events. 
\item A preliminary evaluation of e/$\gamma$ and e/$\pi^0$
discrimination capability by means of the specific ionization measured
on the wires at the beginning of the candidate track. This measurement
is limited by the size of the chamber.
\end{itemize}

The data collected in 1997 offer the unique opportunity to study 
nuclear effects in the Argon nucleus and to assess the identification and
reconstruction capability of a LAr TPC for low-multiplicity neutrino events. 
The present work shows, for the first time, a comprehensive set of results 
from the 1997 test (see \cite{ref:pietropaolo,ref:tesi_alessandro,ref:tesi,ref:curioni} 
for preliminary results). The structure of the paper is as follows: the
experimental setup and the instrumental performance of the 50 liters TPC
are discussed in Sec.~\ref{sec:Setup} and Sec.~\ref{sec:TPC},
respectively. In Sec.~\ref{sec:event_reconstruction} the event
reconstruction and particle identification are detailed. The analysis
of a ``golden-sample'' of quasi-elastic (QE) $\nu_\mu$ CC interactions is presented in
Sec.~\ref{sec:Analysis_QE} together with a comparison with theoretical
expectations.

\section{The experimental setup}
\label{sec:Setup}

\begin{figure*}[htbp]
\centering
\epsfig{file=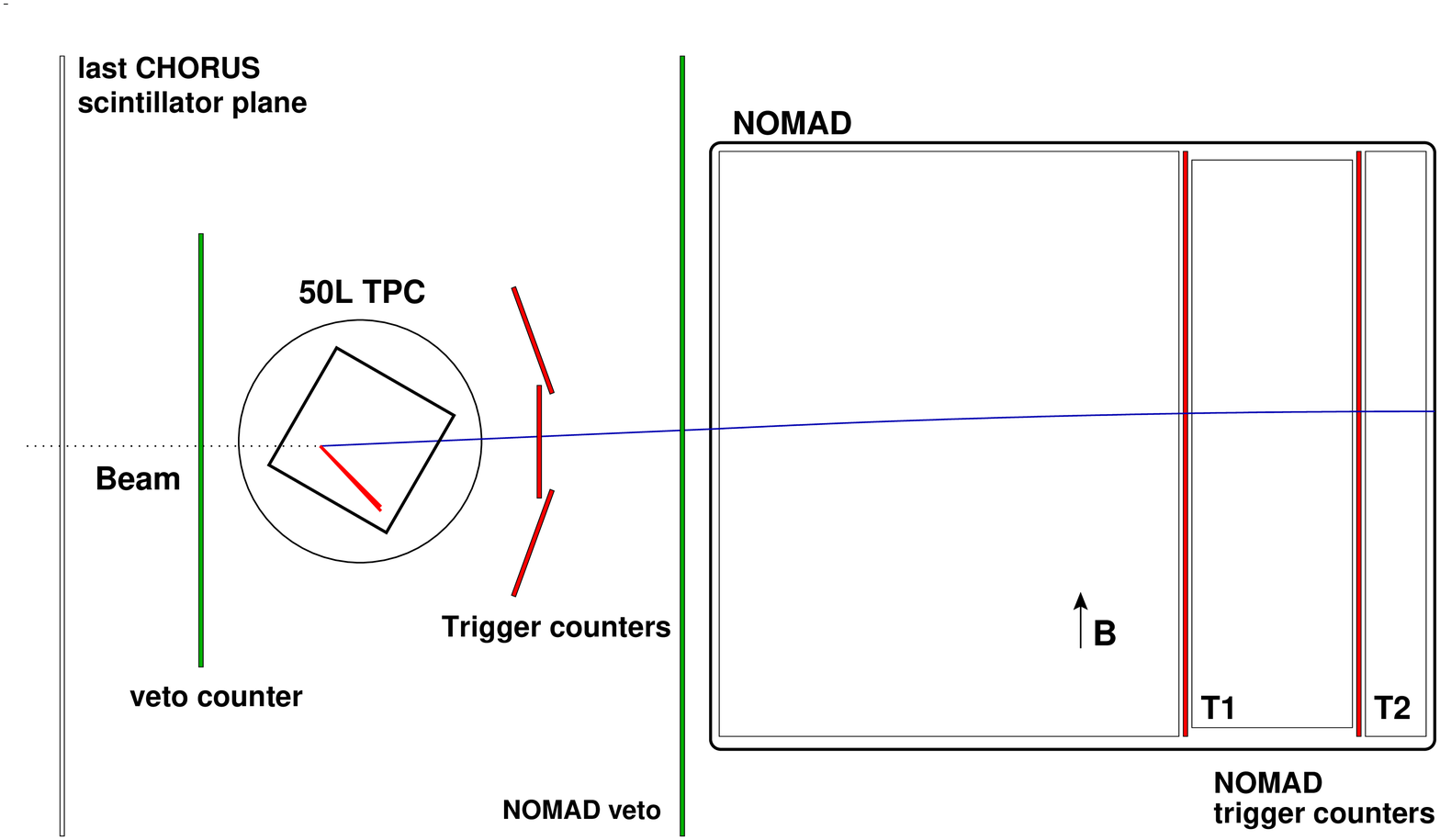,width=14cm}
\caption{A sketch of the experimental setup (top view). Relative sizes are not
to scale.}
\label{fig:setup}
\end{figure*}

The LAr TPC was placed on a platform 4.5 meters above
ground, right in between the CHORUS and NOMAD detectors
(Fig.~\ref{fig:setup}). The modest size of the LAr TPC fiducial
volume ($\sim$50 liters), coupled with the high energy of the WANF
$\nu$ beam, made necessary a muon spectrometer downstream the TPC.  A
coincidence with the NOMAD DAQ was set up to use the detectors located
into the NOMAD magnetic dipole as a magnetic spectrometer. The
experimental setup was completed with additional counters for the
trigger and veto systems.

\begin{figure*}[htbp]
\begin{center}
  \begin {tabular}{cc}
    \epsfysize=8.cm\epsfxsize=6.cm\epsffile{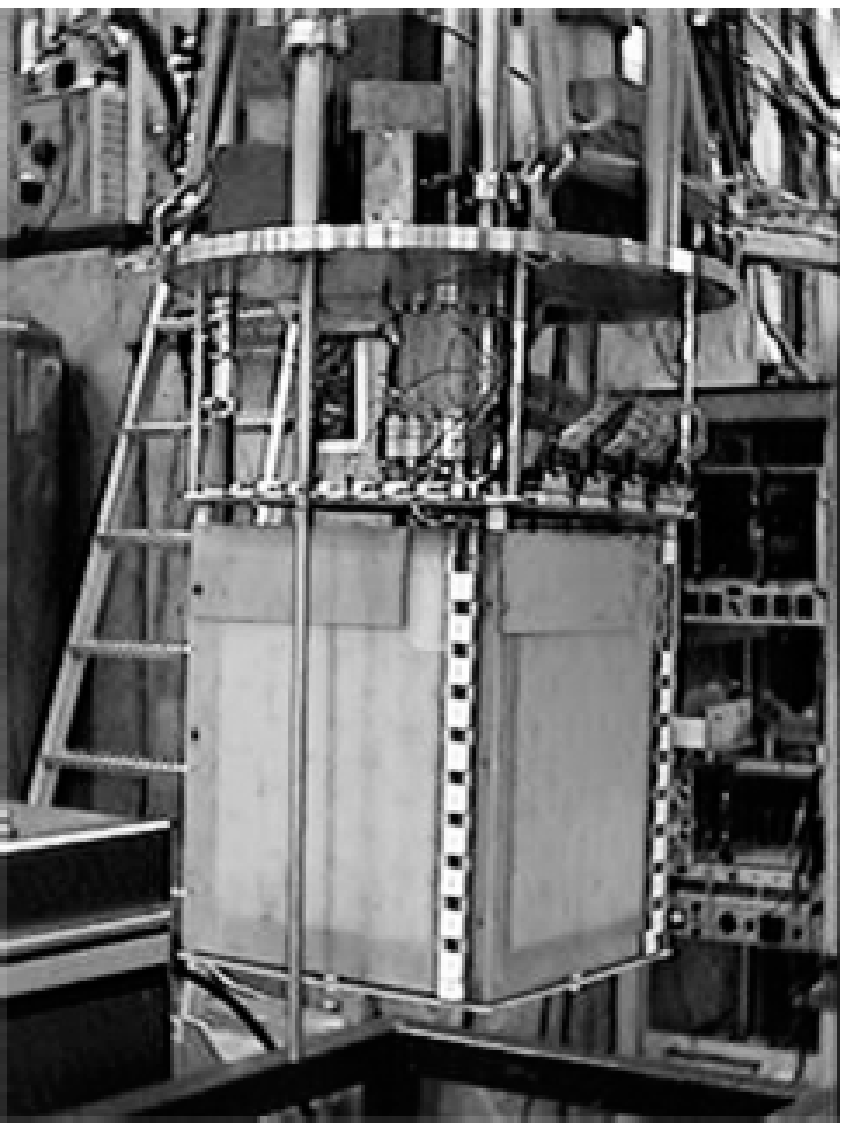} &
    \epsfysize=8.cm\epsfxsize=6.cm\epsffile{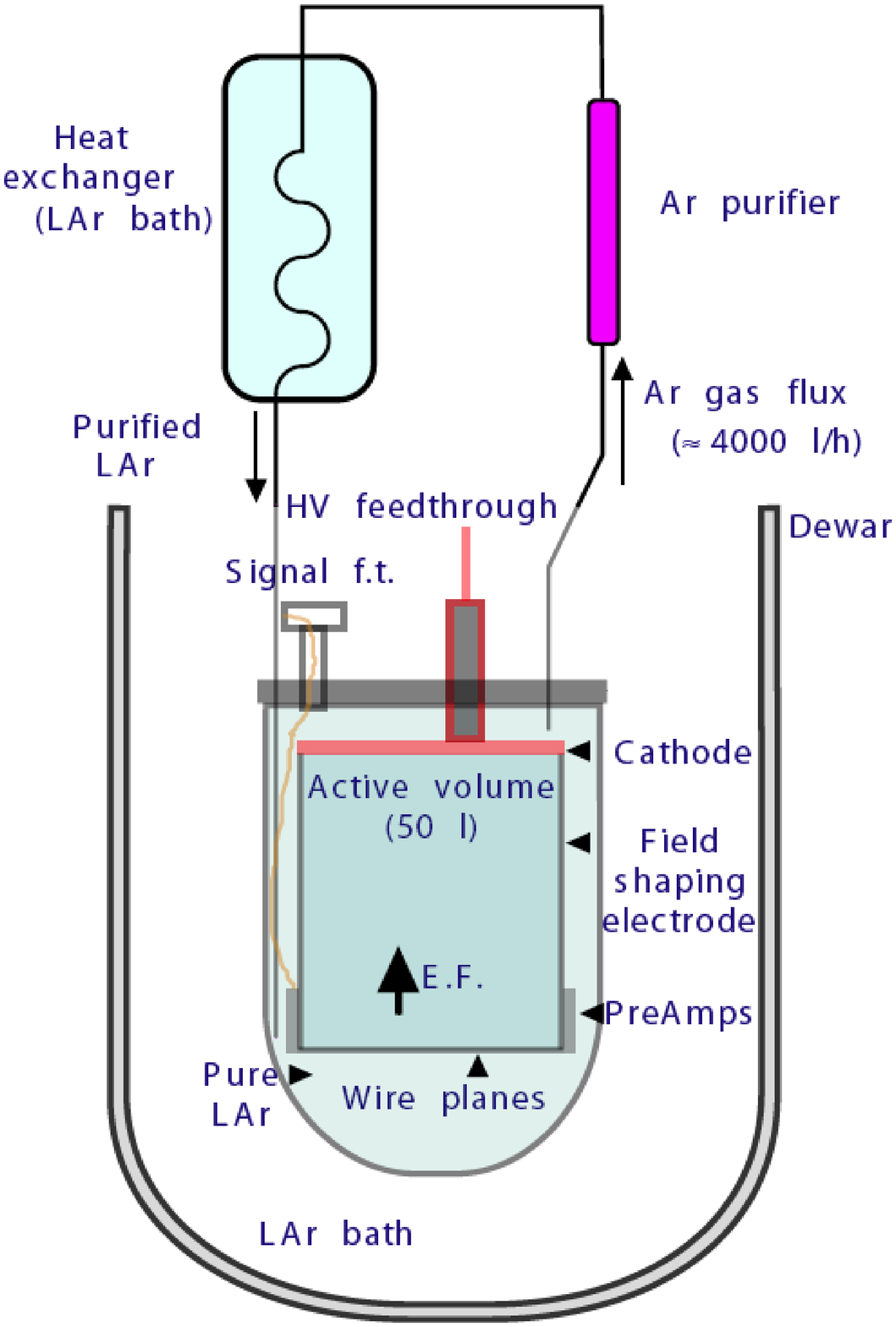} 
  \end{tabular}	
\caption{The 50 liters Liquid Argon Time Projection Chamber.}	
\label{fig:50L_schema}
\end{center}
\end{figure*}

The LAr TPC (Fig.\ref{fig:50L_schema}) has an active volume of
32$\times$32$\times$46.8~cm$^3$, enclosed in a stainless steel vessel
in the shape of a bowed-bottom cylinder 90 cm high and with a radius
of 35 cm. The active volume corresponds to 67 kg of Liquid
Argon ($T=87 $K at 1 atm, $\rho=1.395$ g/cm$^3$). 
Ionization electrons produced by the passage of charged
particles drift vertically toward the anode by
means of a constant electric field of 214~V/cm. The readout area
(anode) is made up of two orthogonal wire planes mounted at a distance
of 4~mm. Each stainless steel wire has a diameter of 100~$\mu$m; the
distance between the wires is 2.54~mm and each plane is made up of 128
wires. The cathode and the lateral field-shaping electrodes are copper
strips (5~mm thickness) positioned on a vetronite support with printed
board techniques. The support was glued on a honeycomb structure
to ensure rigidity. The distance between two adjacent strips is
10~mm. The 10~KV drift voltage is distributed to the strips by means
of 100~M$\Omega$ resistors. The ionization electrons drift through the
first wire plane (induction plane) so that the integrated signal
induced in it is zero. The second plane (collection plane) stops the
electrons and acts as charge collector.  The mean charge per unit
pitch (2.54~mm) for a mip crossing the chamber horizontally
corresponds to about $2.3 \times 10^4$ electrons ($\sim$4 fC) assuming
no charge loss along the drift volume.  The smallness of the signal
and the absence of an amplification phase requires very low noise
charge preamplifiers.  They were based on JFET transistors and have
been operated into the liquefied gas in the
proximity of the readout wires. The polarization voltage (15~V) is
distributed separately to each preamplifier to minimize the number of
dead channels in case of failure of one of the components during data
taking. The output signal is transmitted up to the amplifiers, which
have been operated in current mode and are located at room
temperature. The amplified signal is brought up to a set of fast 8-bit
ADC.  This configuration~\cite{ref:pietropaolo}
allowed a signal to noise ratio of 11 with mip  signals equivalent
to 10 ADC counts and no saturation even in the occurrence of
e.m. showers. The signal is sampled with a 2.5~MHz frequency for a
duration of 500~$\mu$s, corresponding to the highest possible drift
time (primary ionization at the cathode). The stream is recorded into
a buffer.  The arrival of a subsequent trigger causes the switch of
the data stream to another buffer (up to 8 buffers are available). The
multi-buffer writing procedure minimize the dead time for signal
recording to less than one sampling time (i.e. 500~$\mu$s).

As mentioned above, the active part of the detector is located inside
a stainless steel cylinder. The connection to the outside area is
obtained through a set of UHV flanges housing the signal, the high
voltage cables and the vacuum feed-through. The cylinder is positioned
into a 1~m diameter dewar partially filled with low purity Liquid
Argon acting as a thermal bath and it is rotated 30$^\circ$ along
the vertical axis with respect to the nominal beam direction to reduce
the number of particles crossing just one readout wire. The ceiling of
the external dewar is in direct contact with air.  The level of Argon
in the dewar is so small that the pure Ar in the inner cylinder can
evaporate as well. The gaseous Ar in the active volume crosses the
feed-through and reaches an Oxisorb filter for
purification~\cite{ref:purification}. After, it is liquefied
inside a second buffer cooled by low-purity Ar and, finally,
recollected into the active volume.  During this test, the Argon has
been doped with Tetramethyl Germanium in order to partially reconvert
scintillation light into ionization \cite{ref:TMG}. 
Continuous recirculation of the Argon through
the purifier keeps the level of contamination stable against
micro-leaks or outgassing. The total Argon consumption necessary to
circulate the pure liquefied gas and to compensate for the heat losses
was $\sim$200 liters per day. 
The performance of the recirculation system 
is described in Sec.~\ref{sec:TPC}.

The chamber has been exposed to the $\nu$ beam produced at the CERN
West Area Neutrino Facility (WANF~\cite{ref:wanf,ref:flux_nomad}). The
primary 450~GeV protons from the CERN SPS are extracted every 14.4~s
in two spills of 6~ms duration each and separated by 2.5~s.  On
average, $1.8 \times 10^{13}$ protons per spill hit a segmented
beryllium target. Secondaries are selected in momentum and focused by
a system of collimators and magnetic lenses. They reach a 289.9~m
decay tunnel followed by an iron and earth absorber.  The experimental
area is located 835~m downstream the target. The mean energy of the
$\nu_\mu$ reaching the detectors is 24.3~GeV when integrating over the active 
NOMAD area ($2.6\times 2.6$ $m^2$), while contaminations from other flavors 
are below 7\% for $\bar{\nu}_\mu$ and $\sim 1$\% for $\nu_e$~\cite{ref:flux_nomad}.
Being the TPC centered on the beam axis and covering a smaller surface, 
an harder neutrino flux is expected (with a mean energy of about 30 GeV).

The trigger is provided by a set of scintillators located between the
chamber and the NOMAD apparatus (Fig.~\ref{fig:setup}). Each of
the 3 trigger scintillator counters has a dimension of
110$\times$27~cm$^2$. They are positioned in a half-circle 60~cm
beyond the center of the chamber.  Incoming charged particle are
vetoed by 5 scintillators mounted in front of the chamber, 50~cm
before its center, and by the last scintillator plane of CHORUS.  The
latter vetoes particles deflected by the CHORUS magnetic field
entering the chamber at large angles with respect to the nominal beam
direction.

The trigger requires the coincidence of the SPS beam spill and at
least one of the trigger scintillators, vetoed by the scintillators
put in front of the chamber and the CHORUS plane. This trigger is put
in coincidence with the two trigger scintillator planes of NOMAD (T1
and T2 in Fig.~\ref{fig:setup})~\cite{ref:trigger_NOMAD}. Moreover, a
trigger is rejected if the NOMAD acquisition system is in BUSY mode or
if the delay with respect to the previous trigger is lower than
500~$\mu$s (``drift protection'')\footnote{This condition inhibits the
occurrence of event overlaps in the multibuffer readout system.}. 
The kinematic reconstruction of the outgoing muon from NOMAD 
(see Sec.\ref{sec:muon_rec})
allowed us to identify the events produced outside the fiducial volume
that trigger the chamber. The vast majority are $\nu$ interactions
in the dewar or in the thermal bath plus a contamination of crossing
muons due to veto inefficiencies. Note that the request of a charged
particle triggering the chamber locally and reaching NOMAD up to T1
and T2 inhibits the acquisitions of neutral-current and $\nu_e$
charged-current events, limiting the sample of the present test to
$\nu_\mu$ charged-current interactions.  The trigger efficiency has
been monitored during data taking through a dedicated sample of
through-going muons. It turned out to be 97\% averaged over the whole
data taking period.

\section{Instrumental performance of the 50 liters TPC}
\label{sec:TPC}

\subsection{Liquid Argon purification and electron lifetime}
\label{sec:electron_lifetime}

A crucial working parameter for a TPC operating with liquefied rare
gases is the lifetime of free electrons in the medium. Primary
electrons produced by the passage of charged particles drift toward
the anode crossing macroscopic distances. In the present case, for an
ionization electron created in the proximity of the cathode, the drift
path-length exceeds 46~cm.  The effectiveness of charge collection at
the readout plane is related to the purity of the Argon since
electron-ion recombination is mainly due to oxygen molecules present
in the LAr bulk~\cite{ref:bettini}. The contamination of
electronegative molecules must be at the level of 0.1~ppb to allow
drifts over ${\cal O(1)}$ meters.  The electron lifetime $\tau_e$ is
defined by:

\begin{equation}
Q(t) \ = \ Q_0 e^{-\frac{t-t_0}{\tau_e}}
\label{equ:exp_law}
\end{equation}

$Q_0$ being the primary deposited ionization charge and $t$ the electron drift
time. The $t_0$ is provided by the trigger scintillation counters.

\begin{figure*}[htbp]
\centering
\epsfig{file=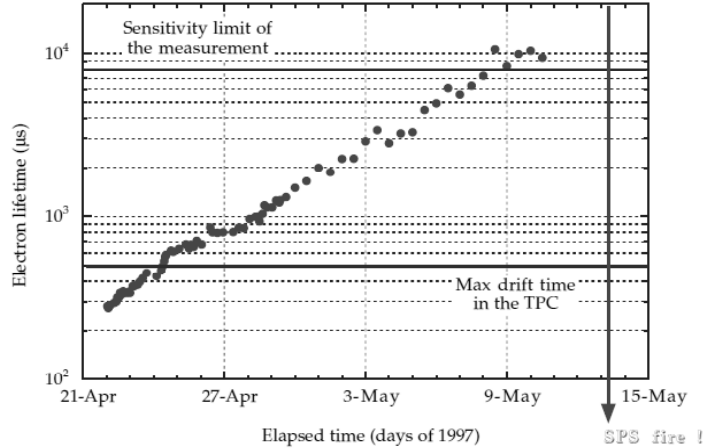,width=10cm}
\caption{Lifetime of the drifting electrons in the 50 liters Liquid Argon TPC 
at the CERN $\nu$ beam. The filling of the chamber with LAr was performed on
the fourth of April 1997. }
\label{fig:eLifetime}
\end{figure*}

The effectiveness of the purification system through direct
measurements of $\tau_e$ has been monitored during data taking by a
dedicated setup first developed by the ICARUS Collaboration in
1989~\cite{ref:monitor_purity}. It consists of a small double-gridded
drift chamber located below the readout planes. Electrons are
generated near the cathode via photoelectric effect driven by a 20~ns
UV laser pulse. Each pulse produces a bunch of 10$^7$ electrons. They
drift toward the anode along the electric field lines and cross a 50
cm drift region between the two transparent grids. The ratio of the
induced signal in the proximity of the grids provides a real time
estimate of $\tau_e$ during data taking and, hence, a measurement of
the Ar contamination (see Fig.~\ref{fig:eLifetime}). 
An independent estimate of $\tau_e$ can be obtained monitoring the
charge attenuation for muons crossing the chamber at various heights. 
Both methods provided consistent results and the electron lifetime averaged
along the whole data taking period turned out to be always higher than 10~ms.
This performance was excellent since the maximum drift time was about 400 $\mu$s
and therefore the attenuation of the ionization over the drift
distance was negligible.

\subsection{Drift velocity}

An absolute determination of the drift velocity and its uniformity
along the fiducial volume of the chamber has been obtained exploiting
the external scintillators and the additional information from NOMAD.
A dedicated trigger selecting through-going muons (``mip sample'')
has been put into operation adding two additional scintillators
(30$\times$30 cm$^2$) before and after the external dewar. The
relative position of the chamber with respect to the NOMAD reference
frame has been obtained by residual minimization of the track
parameters recorded both by the TPC and the NOMAD data acquisition
system~\cite{ref:tesi}.  The drift velocity is obtained by fitting the
absolute vertical position of the muon as reconstructed by NOMAD
trackers versus the drift time measured locally by the TPC. The
corresponding drift velocity is $v_d = 0.905 \pm 0.005$ mm/$\mu$s 
(this value differs from the one quoted in~\cite{ref:purity} due to 
different running conditions of pressure and temperature). No
systematic biases have been observed in the fiducial volume used for
the neutrino interaction analysis (Sec.\ref{sec:Analysis_QE}).

\section{Event reconstruction and particle identification}
\label{sec:event_reconstruction}

\subsection{Spatial reconstruction}
\label{sec:sparec}

\begin{figure*}[htbp]
\centering
\epsfig{file=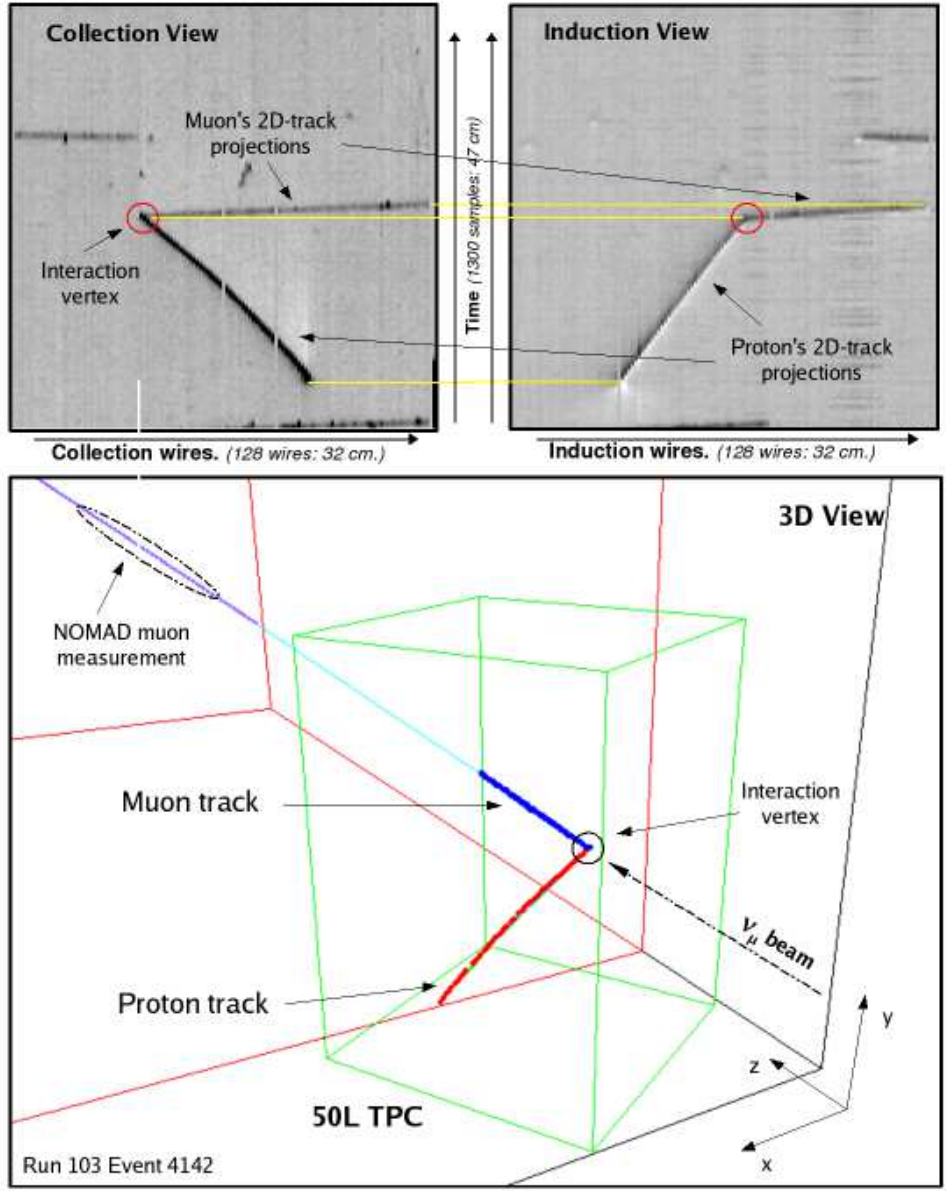,width=14cm}
\caption{Example of the 3D reconstruction of a low-multiplicity $\nu_{\mu}$ CC event.
The raw image from collection and induction wire planes (on top):
Hits and 2D track projections have been identified. 
Three dimensional view of the reconstructed event (bottom picture) embedded in a 
3D recreation of the experimental setup described in Sec.~\ref{sec:Setup}. }
\label{fig:Golden_rec}
\end{figure*}

\begin{figure*}[htbp]
\centering
\epsfig{file=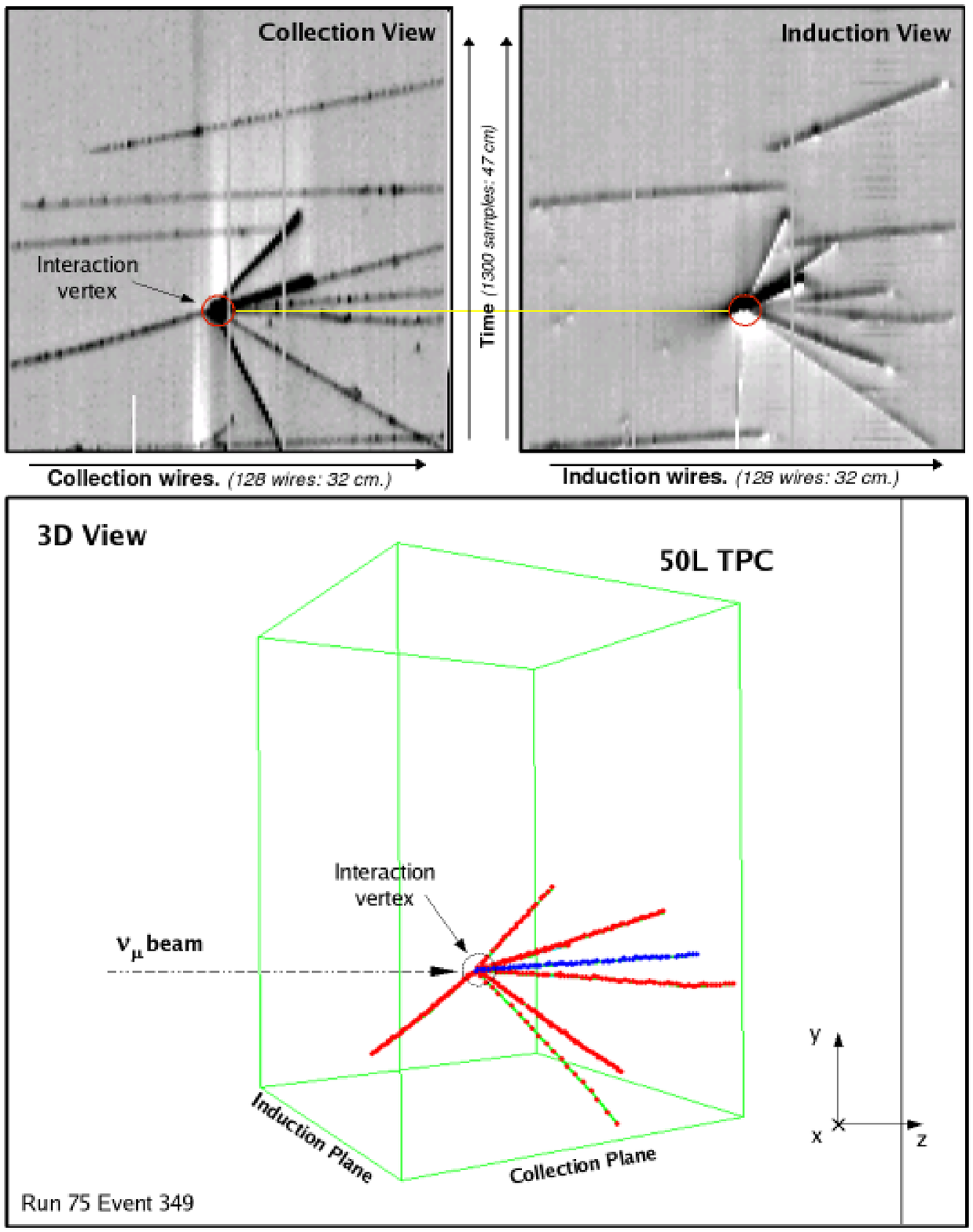,width=14cm}
\caption{(Top) The raw images of a high multiplicity event ($\nu_\mu$ CC DIS) 
in the collection (left panel) and induction plane (right panel). Below, 
the 3D reconstruction  of a high multiplicity event ($\nu_\mu$ CC DIS) with 
eight primary particles. Two of them stop in LAr and are recognized as protons.}
\label{fig:DIS_rec}
\end{figure*}

Interactions happening in the TPC fiducial volume were fully registered in
two 2D images with a common coordinate (time), with full calorimetric
information associated to each point.
Each 2D-image represents the signal amplitude digitized by the ADC (in
a linear gray scale) versus the time sample (drift coordinate) and the
wire number.  Fig.~\ref{fig:Golden_rec} and Fig.~\ref{fig:DIS_rec} (top part)
show the raw images of two $\nu_\mu$ CC events in the collection (left panel) 
and induction plane (right panel). 
These raw images represent the time evolution (in vertical axis) 
of the signal induced in the wires (in horizontal axis).

The purpose of the reconstruction procedure is to extract the physical information 
provided by the wire output signals, i.e. the energy deposited by the different 
particles and the point where such a deposition has occurred, to build a complete 3D
and calorimetric picture of the event. In this sense, the spatial reconstruction of 
different tracks is needed first in order to compute the calibration factors 
entering in the calorimetric reconstruction of the events, as explained in 
Sec.~\ref{sec:calrec}. A detailed description of the spatial reconstruction tools 
was reported elsewhere \cite{ref:600ton,ref:ricotesi}. For the present analyses
new significant improvements have been carried out \cite{ref:delaossatesi}.

The basic building block of a track is called a hit, defined as the segment of track 
whose energy is detected by a given wire. 
A precise determination of the hit position and charge is carried out 
fitting the wire signal to an analytical function which describes the detector response 
\cite{ref:600ton}. Identified nearby hits are associated into 2-dimensional 
clusters. Track candidates are searched for by means of a tree algorithm 
as in \cite{ref:TreeFin} and/or using a neural network-based algorithm as 
in \cite{ref:NNFin}. Both methods efficiently detect interaction vertexes and finally,
split clusters into smooth 2D tracks. Finally, 3D tracks are built matching 2D track
projections from both views.


In Fig.~\ref{fig:Golden_rec}, we show an example of the full reconstruction procedure
for a low-multiplicity $\nu_{\mu}$ CC event. This algorithm allows the reconstruction
of more complicated event topologies (see Fig.~\ref{fig:DIS_rec}). 
This is the first time that interactions of multi-GeV accelerator neutrinos occurring 
in a LAr TPC are fully reconstructed in 3D. 

\subsection{Calorimetric reconstruction}
\label{sec:calrec}

The ionization charge is precisely measured at the collection wire plane. The energy $E$
associated to a given hit is:
\begin{equation}
E \ = \ \frac{CW}{R} \ Q(t)
\label{equ:QtoE}
\end{equation}
where $C$ is the calibration factor; $W$ is the average energy needed for the 
creation of an electron-ion pair (i.e. 23.6 eV); $R$ is the electron-ion recombination factor
and $Q(t)$ is the corrected charge by the electron lifetime from (\ref{equ:exp_law})
in Sec.~\ref{sec:electron_lifetime}.

\begin{figure*}[htbp]
\centering
\epsfig{file=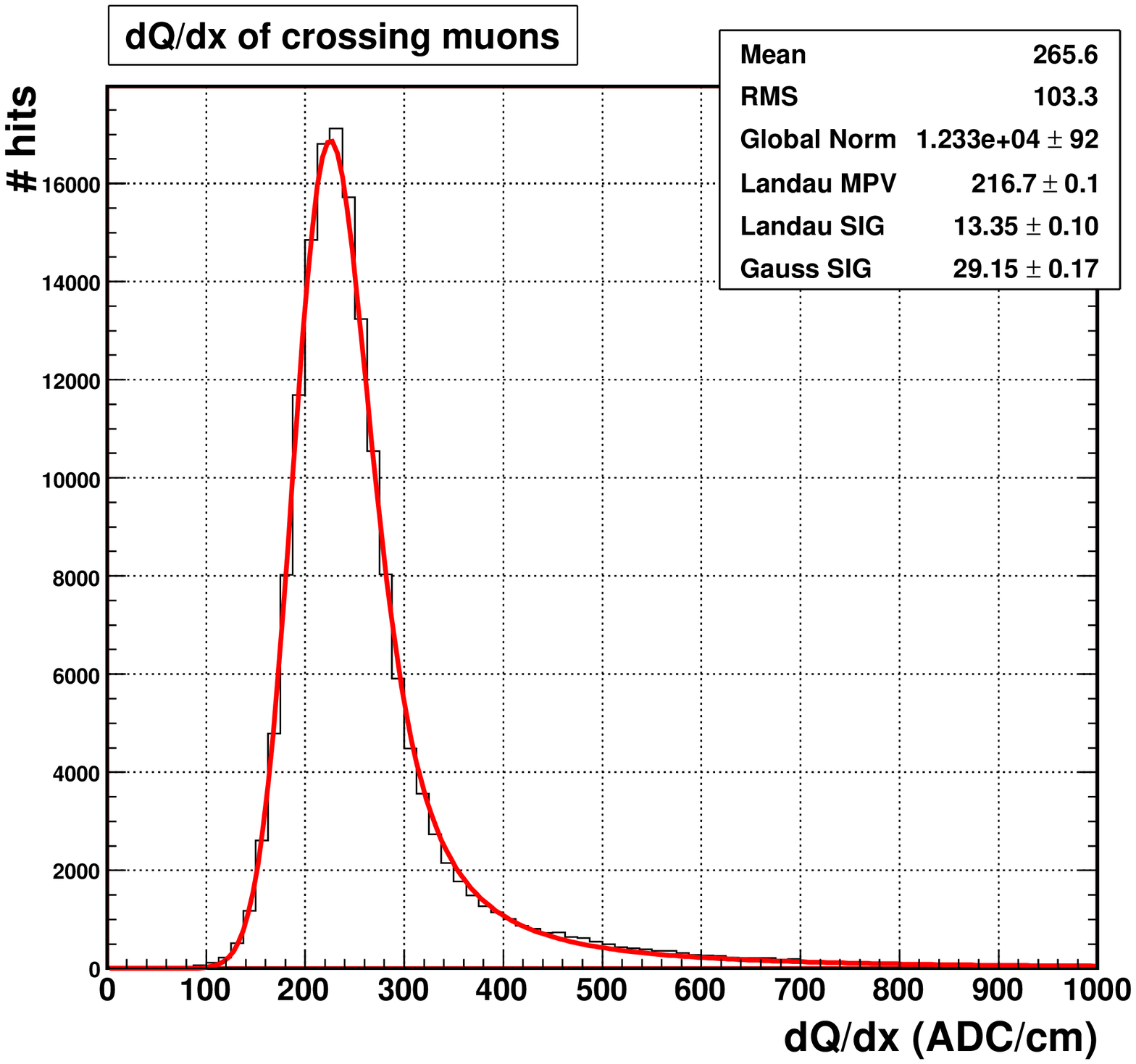,width=10cm}
\caption{$dQ/dx$ of crossing muons. The distribution is fitted to a convoluted 
Landau-Gaussian distribution.}
\label{fig:dqdx_muons}
\end{figure*}

The calibration factor $C$ converts the detector measuring units (ADC counts) into charge
units (fC). To measure the overall calibration factor $\alpha \equiv \frac{CW}{R}$ 
in (\ref{equ:QtoE}), we have collected a sample of around 3000 through-going muons coming
from $\nu_\mu$ CC interactions.
The deposited charge per unit of length ($dQ/dx$) is precisely measured 
(see Fig.~\ref{fig:dqdx_muons}) and fitted to a convoluted
Landau-Gaussian function to obtain the most probable
energy loss for mips in terms of detector charge units: 
$\left\langle dQ/dx \right\rangle_{mip} = 216.7 \pm 0.1$ ADC counts.
The most probable energy loss is a better defined quantity than the average value 
in the case of a Landau distribution, moreover it slightly depends on 
the energy in the range of the muons being considered.
From this measurement, $\alpha$ can be obtained assuming that according to
the theoretical prediction for muons in the range of energies of the beam, the most probable
energy loss $\left\langle dE/dx \right\rangle_{mip} = 1.736 \pm 0.002$ MeV/cm:
\begin{equation}
\alpha_{mip} \ \equiv \ \frac{CW}{R_{mip}} \ = \ 
\frac{\left\langle dE/dx \right\rangle_{mip}}{\left\langle dQ/dx \right\rangle_{mip}}
\ = \ 8.01 \pm 0.01 \times 10^{-3} \ MeV/ADC
\label{equ:cal_factor}
\end{equation}

The recombination factor $R$ in (\ref{equ:QtoE}) 
depends on the absorber medium, on the applied electric field and on the density
of released charge, i.e. on $dE/dx$. This is the reason why we take in (\ref{equ:cal_factor}) 
$R_{mip}$ (its value for mips).
The dependence of $R$ with $dE/dx$ can be modeled by Birk's law
\cite{ref:Birks} which has been successfully applied
to LAr detectors in \cite{ref:Rec_factor}:
\begin{equation}
\frac{dQ}{dx} \ = \ \frac{ a \ dE/dx }{ 1\ +\ k_B \ dE/dx }
\label{equ:birks}
\end{equation}
$dE/dx$ being the pure Bethe-Bloch ionization loss, 
$a$ the energy to charge conversion factor and $k_B$ the Birks
coefficient accounting for quenching due to recombination.

\begin{figure*}[htbp]
\begin{center}
\mbox{\epsfig{file=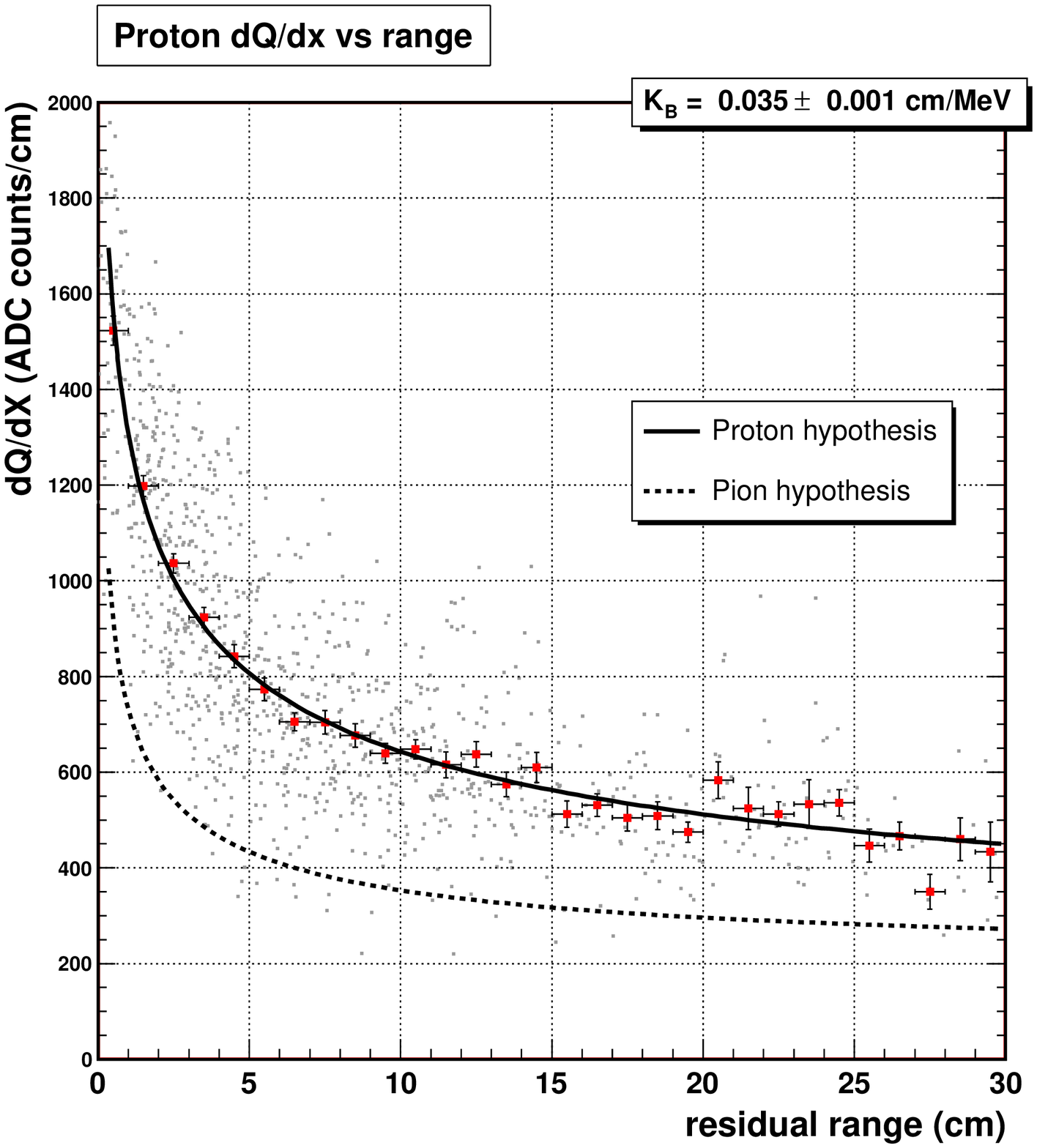,width=10cm}}
\end{center}
\caption{$dQ/dx$ as a function of the residual range for protons.
Dots are direct measurements from the reconstructed hits of the proton tracks, while
the points with errors bars are the estimated most probable of $dQ/dx$ for each bin 
of the residual range. Data is fitted with a Bethe-Bloch corrected by the detector 
response function (Birk's law) in the hypothesis of pion (dashed line) and proton (full
line)}
\label{fig:dqdx}
\end{figure*}

The size of the range of the fully contained protons collected in the ``golden sample'' 
considered in Sec.~\ref{sec:Analysis_QE},
offers the opportunity to precisely test the energy loss pattern in the active
medium. Fig.~\ref{fig:dqdx} shows the observed charge per unit length
($dQ/dx$) versus the residual range for the protons\footnote{The
residual range is the actual range minus the range already covered at
the time of the deposition of a given $dQ/dx$.}. Dots are direct measurements
from the reconstructed hits of the proton tracks. From this distribution 
we estimate the most probable value of $dQ/dx$ for each bin of the residual range. 
Finally, from this data we fit $k_B~= 0.035 \pm 0.001$ cm/MeV in
(\ref{equ:birks}) for the proton hypothesis (continuous line).

Based on the knowledge of the calibration parameters, a precise
determination of the deposited energy can be obtained through direct charge 
measurements by means of (\ref{equ:birks}). Moreover, pure Bethe-Bloch ionization loss
can be corrected by the detector response in the region of high $dQ/dx$, where quenching
effects are more sizable. This correction is important for particle identification
in LAr, which exploits the different energy loss of particles near the stopping point.

\subsection{Proton reconstruction}
\label{sec:prec}

\begin{figure*}[htbp]
\begin{center}
  \begin {tabular}{cc}
    \epsfysize=7.cm\epsfxsize=6.8cm\epsffile{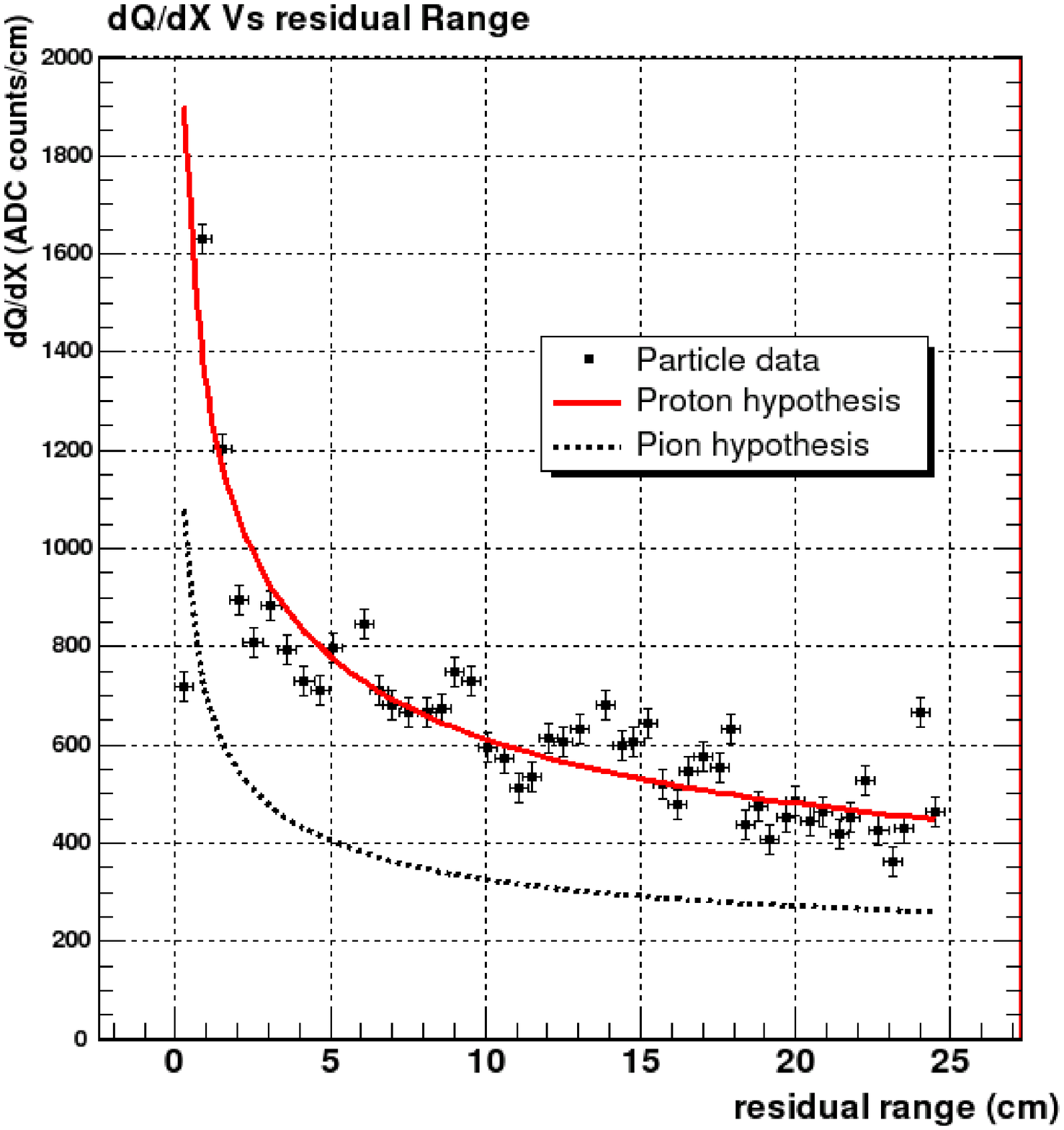} &
    \epsfysize=7.cm\epsfxsize=6.8cm\epsffile{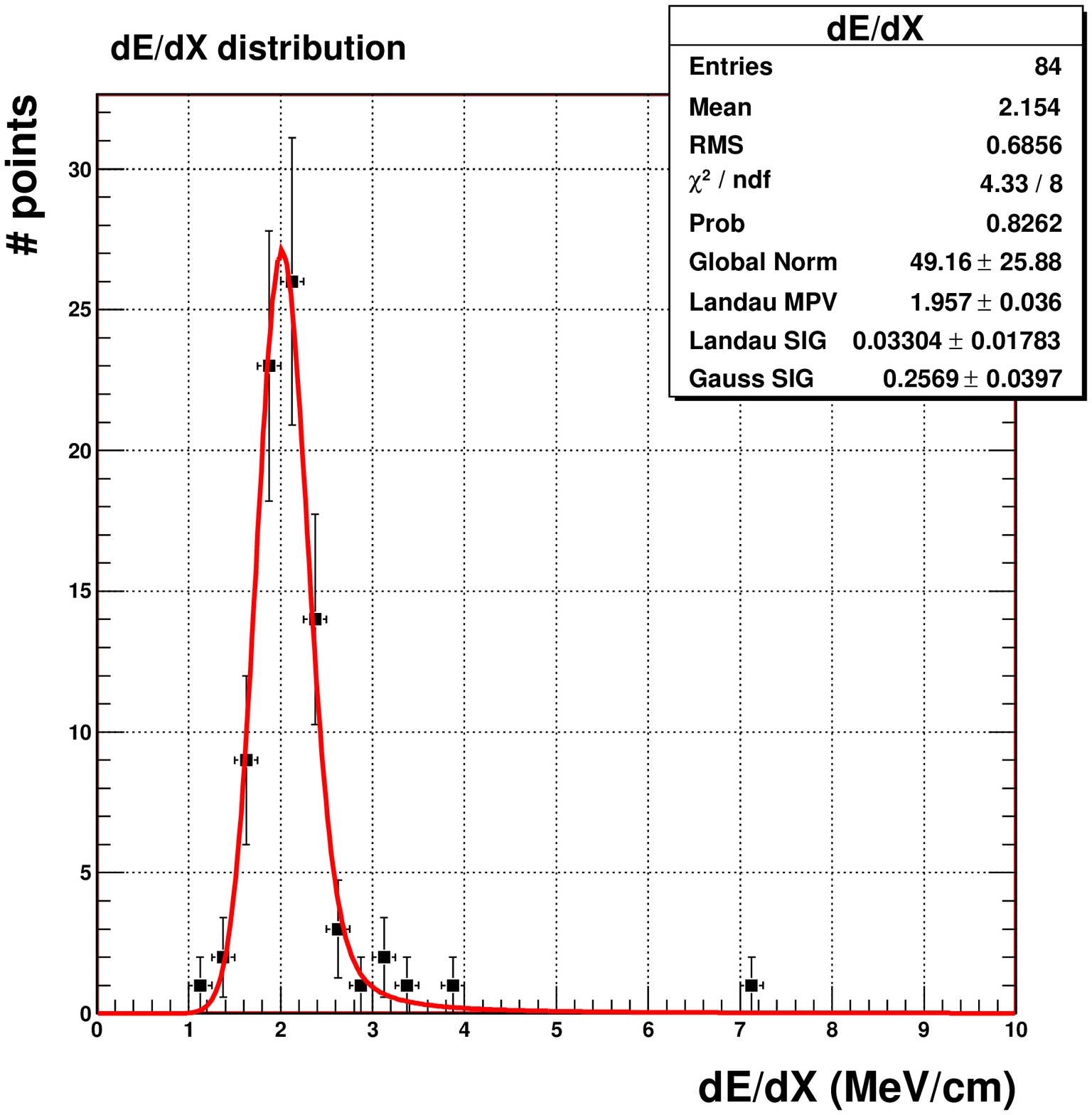} 
  \end{tabular}	
\caption{(left) $dQ/dx$ as a function of the residual range for the
reconstructed proton in the event shown in Fig.~\ref{fig:Golden_rec}: continuous (dashed)
line represents the expected behavior of protons (pions) in the detector. 
(right) $dE/dx$ distribution for the muon reconstructed in 
Fig.~\ref{fig:Golden_rec}: data is fitted to a convoluted Landau-Gaussian distribution.}
\label{fig:dqdx_evt}	
\end{center}
\end{figure*}

Proton identification and momentum measurement were
performed using only information provided by the TPC. The discrimination
between protons and charged pions is performed exploiting their
different energy loss behavior as a function of the range.
For the case of candidates stopping in the fiducial volume of the chamber,
as the ones of the ``golden sample'', p/$\pi^\pm$ separation is
unique (see Fig.~\ref{fig:dqdx_evt} [left]): $dQ/dx$ is measured along 
the proton candidate track and is compared with the different particle
hypothesis. Monte-Carlo studies reveal that almost 100\% of protons are 
identified as such on the basis of their $dQ/dx$ shape on the vicinity of the
stopping point; in addition, the fraction of pions and kaons misidentified
as protons is negligible.


Once we identify a contained proton, its kinetic energy is calculated
from range, which only depends on the spatial reconstruction of the tracks.
In this case, the momentum uncertainty is dominated by 
the finite pitch of the wires (2.54 mm)\footnote{The equivalent pitch in the
vertical direction (drift direction) is much smaller due to the high
sampling rate of the fast ADC ($360\ \mu$m)} and the performance of the
reconstruction tools. These have been tested using a Monte-Carlo simulation
of the detector. The resolution on the kinetic energy ($T_p$) varies from
3.3\% for protons with $T_p=50$ MeV up to 1\% for $T_p$ larger than
200 MeV. The angular resolution in the collection or induction view
depends on the number of wires $N$ hit by the particle along its path;
it is $\sigma \simeq 0.36\sqrt{12}/(2.54\ N^{3/2})$, corresponding
e.g. to 15 mrad for N=10.


\subsection{Muon reconstruction}
\label{sec:muon_rec}

The kinematic reconstruction of the outgoing muons exploits the
tracking capability of NOMAD. An event triggering the chamber will
have at least one penetrating track reaching the T1 and T2 trigger
scintillators bracketing the TRD of
NOMAD~\cite{ref:trigger_NOMAD}. The corresponding track, nearly
horizontal at the entrance of the NOMAD drift chamber volume, is
reconstructed with an average momentum precision of $\sigma_p/p \sim
0.05/\sqrt{L} \oplus 0.008p/L^{5/2}$, $L$ being the visible range in
the volume itself expressed in meters and $p$ the particle momentum in
GeV. 
A 10~GeV horizontal muon crossing all the chambers
($L\sim 5m$) is reconstructed with a precision of 2.2\%.  The
reconstructed particle is traced back to the TPC accounting for the
magnetic field and the presence of the forward NOMAD calorimeter.  The
latter introduces an additional - in fact, dominant - source of
uncertainty due to multiple scattering (MS) in iron (190~cm for a
horizontal muon). For small scattering angles ($\theta \ll 1$~rad), the
MS uncertainty on the transverse momentum is independent of $p$ and it
turns out to be $\sim 140$~MeV. The correctness of the back-tracing
procedure has been cross checked comparing the direction angles of the
particles belonging to the mip sample, as measured by the TPC, with
the corresponding quantity from NOMAD\footnote{Due to the high
sampling rate of the TPC, the angular resolution in the $y-z$ plane,
i.e. in the vertical plane along the nominal beam direction
($z$-axis), is dominated by the NOMAD uncertainty.}. Due to the
large amount of material between the TPC and the NOMAD trackers (more
than 11 interaction lengths) and the small distance between the
chamber and the calorimeter, the $\pi/\mu$ misidentification
probability is negligible for the present study, even accounting for
possible $\pi$ decays in flight. 

\section{Analysis of quasi elastic $\nu_\mu$ interaction} 
\label{sec:Analysis_QE}

\subsection{Data taking and event selection}
\label{sec:Sel_QE}

The data set recorded with the 50L chamber amounts to
$1.2\times 10^{19}$ protons on target. 
As mentioned in Sec.\ref{sec:Setup}, the trigger efficiency was monitored
during data taking and its integrated value is 97\%. Additional losses
of statistics are due to the TPC (3\%) and NOMAD (15\%) dead time and
to detector faults. These contributions add up to give an effective
lifetime of 75\%. Over the whole data taking period around 70000 triggers 
were collected in the TPC from which 20000 have at least 
a reconstructed muon possibly transversing the fiducial volume 
(see Sec.~\ref{sec:muon_rec}). From all these a priory $\nu_\mu$ CC candidates,
a visual scanning reveals that around half of them show a vertex in the 
fiducial volume, the rest are $\nu$ interactions in the surrounding LAr bath 
(which also give trigger) plus a contamination of crossing muons due to veto 
inefficiencies. Therefore, we have a collection of around 10000 $\nu_\mu$ CC events
from which we have selected a set of 86 events called the ``golden sample''.
This set consists of events with an identified proton 
of kinetic energy larger than 40~MeV fully contained in the TPC 
and one muon whose direction extrapolated from NOMAD matches the outgoing 
track in the TPC (see Sec.\ref{sec:event_reconstruction}).  
The distance of the primary interaction vertex to any of the
TPC walls has to be greater than 1~cm. The muon candidate track
projected onto the wire plane must be longer than 12 wire pitches. The
event is accepted even in the presence of other stopping particles, as
far as their visible range does not exceed the range of a proton of 40
MeV kinetic energy. If other tracks than the
identified muon leave the TPC or at least one converted photon with
energy greater than 10~MeV is present in the fiducial volume, the
event is rejected. The tightness of these selections makes the ``golden sample'' a very
clean topology for visual scanning (Fig.~\ref{fig:Golden_rec}).

The request of containment is very severe since the chamber volume is small.
For lower $T_p$ values, the proton range is comparable with
the wire pitch and neither the proton momentum nor the interaction
vertex can be reconstructed with due precision. However, for
$T_p>40$ MeV the $\pi^{\pm}$/p misidentification probability is
negligible (see Sec.~\ref{sec:prec}). The ``golden sample'' contains pure 
QE interactions ($\nu_\mu \ n \ \rightarrow \mu^- \ p$) plus 
an intrinsic background dominated by resonant production followed by pion 
absorption in the nucleus ( $\nu_\mu \ p \
\rightarrow \Delta^{++} \mu^- \rightarrow \mu^- \ p \ \pi^+$, 
$\nu_\mu \ n \ \rightarrow \Delta^{+} \mu^- \rightarrow
\mu^- \ p \ \pi^0$). There is also an instrumental background due to final
state neutral particles ($\pi^0$'s, $n$'s and $\gamma$'s) that scape undetected.
On the other hand, the $\nu_\mu \ n \ \rightarrow \Delta^{+} \mu^- \rightarrow \mu^- \ n \ \pi^+$
contamination is negligible in this tightly selected sample due to the superb
$\pi^{\pm}$/p identification capabilities of the LAr TPC.

\subsection{The Monte-Carlo data sample}
\label{sec:montecarlo}

\begin{figure*}[htbp]
\centering
\epsfig{file=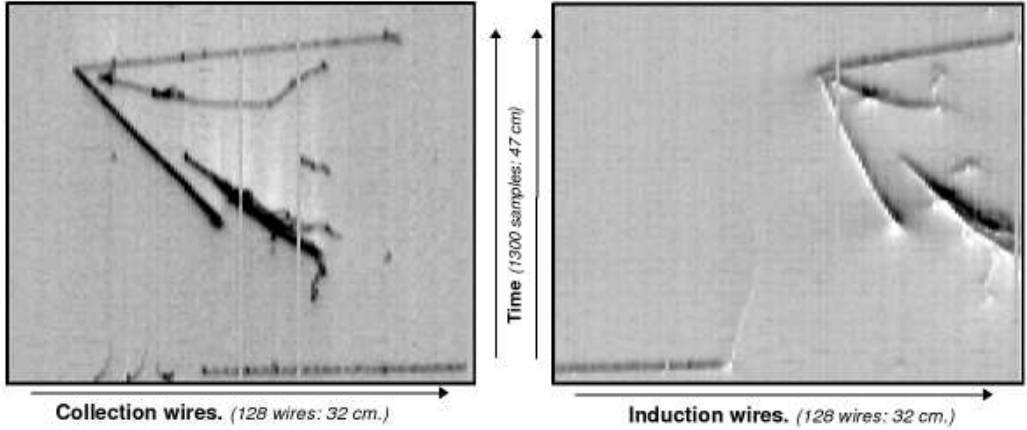,width=14cm}
\caption{The raw image of a low multiplicity event in the
collection (left) and induction plane (right). The event
is reconstructed as ($\nu_\mu \ n \rightarrow \mu^- \Delta^+
\rightarrow \mu^- \ p \ \pi^0$) with a mip leaving the chamber, an
identified stopping proton and a pair of converted photons from the
$\pi^0$ decay. When these photons escape from the chamber, the event is 
tagged as a ``golden event''.}
\label{fig:resonance}
\end{figure*}

The selection efficiency for QE interactions and their
intrinsic background have been evaluated using a Monte-Carlo sample
generated using the FLUKA package~\cite{ref:FLUKA, ref:FLUKATWO}, 
which offers a full description
of nuclear effects in neutrino interactions. About 18000 $\nu_\mu$ CC
QE events have been generated simulating the WANF beam for the 50L chamber
geometrical acceptance and final state particles were tracked in LAr
using the GEANT4 package \cite{ref:geant}. The energy deposition of all final 
particles is digitized emulating the TPC wire read-out, in order to 
reproduce detector resolutions and offline reconstruction efficiencies
(see Sec.~\ref{sec:event_reconstruction}).
We saw that 16~\% of the whole sample of generated QE interactions belong
to the ``golden sample''.
On the other hand, we have analyzed a sample of $10^5$
$\nu_\mu$ CC deep-inelastic (DIS) and resonant (RES) events in order to evaluate
inefficiencies of the vetoing selections for
$\nu_\mu~n~\rightarrow \Delta^{+} \mu^- \rightarrow \mu^-~p~\pi^0$ 
(i.e. the probability to miss both the decay photons of the
$\pi^0 \rightarrow \gamma \gamma$ or the $e^+e^- \gamma$ system in
case of $\pi^0$ Dalitz decay) and for events with several neutral particles
(neutrons and energetic photons from nuclear interactions) which scape from the detector
(see Fig.~\ref{fig:resonance}).
The search for this irreducible background in the DIS plus RES samples reveals 
that 0.14\% are ``golden-like'' events.

Taking into account the relative weight between QE and
non-QE events\footnote{Respect to the total number of $\nu_\mu~CC$ events, 
2.3\% are QE, 91.5\% are DIS and the remaining 6.2\% corresponds to resonances events. 
The simulation of the latest is not done treating each resonance individually,
an average of them is taken into account instead.},
we find that the total expected contamination of the ``golden sample'' 
is 20\% (around 40\% of them correspond to events with an escaping
$\pi^0$). The Monte-Carlo prediction for escaping $\pi^0$
can be checked using test samples that consist of golden events with one ($N_1$) or
two ($N_2$) converted photons pointing to the interaction
vertex. Assuming the gamma identification probability to be
uncorrelated for the two photons, we have
$N_1=2N(1-\epsilon_\gamma)\epsilon_\gamma$ and
$N_2=N\epsilon_\gamma^2$, $N$ being the (unknown) overall rate of $p \
\mu^- \ \pi^0$ final states and $\epsilon_\gamma$ the photon
identification efficiency.  After the scanning of the test samples,
$\epsilon_\gamma = (N_1/2N_2 +1)^{-1}$ turned out to be (43$\pm$ 9) \%.
Hence, the probability of missing both gammas is (32$\pm$10) \% to
be compared with the MC calculation of 20.4\%. 

\subsection{Event rates}
\label{sec:Ev_rates}

We use Monte-Carlo to estimate the expected number of
events in the detector. The simulation of the beam predicts a flux of 
$2.37 \times 10^{-7} \nu_\mu$CC/$cm^2$/p.o.t.
over the TPC exposing area. This flux convoluted with the neutrino cross sections 
and scaled to the fiducial mass of the detector gives an event rate of 
$2.05 \times 10^{-15} \nu_\mu$CC/p.o.t.
Now assuming a total exposure of $1.2\times 10^{19}$ p.o.t. and an effective lifetime 
of 75\% (see Sec.~\ref{sec:Sel_QE}), the total number of $\nu_\mu$ CC is equal 
to 18450. Out of this 18450 events, only muons above a certain momentum threshold and 
within angular acceptance will trigger the detector.
Taking the real data muons that triggered the system, we saw that most of 
them are above 8 GeV in momentum and below 300 mrad in angle. 
If we apply those cuts to the Monte-Carlo samples, we end up
with 18450 $\times$ 0.023 (QE fraction) $\times$ 0.95 (efficiency 
of the acceptance cuts for QE), equal to 400 quasi-elastic events.
For DIS+RES events: 18450 $\times$ 0.977 (fraction of DIS+RES) $\times$ 0.65 (efficiency 
of acceptance cuts) gives 11700.
In total we expect 12100 events to be compared with the 10000 $\nu_\mu$ CC we have 
after visual scanning of {\it good triggers} (see Sec.~\ref{sec:Sel_QE}). 

The expected number of ``golden events'' is obtained taking the 16\% (golden fraction)
of the total QE, and adding the corresponding 20\% due to background 
contribution (see Sec.~\ref{sec:montecarlo}), which finally gives $80\pm9(stat)\pm13(sys)$
``golden events'' to be compared with the 86 we observe. The systematic uncertainty
is dominated by the fraction of QE events (2.3\%) and the beam simulation (8\%).

\subsection{Analysis of quasi-elastic interactions}

\begin{figure*}[htbp]
\begin{center}
  \begin {tabular}{cc}
    \epsfysize=7.cm\epsfxsize=6.8cm\epsffile{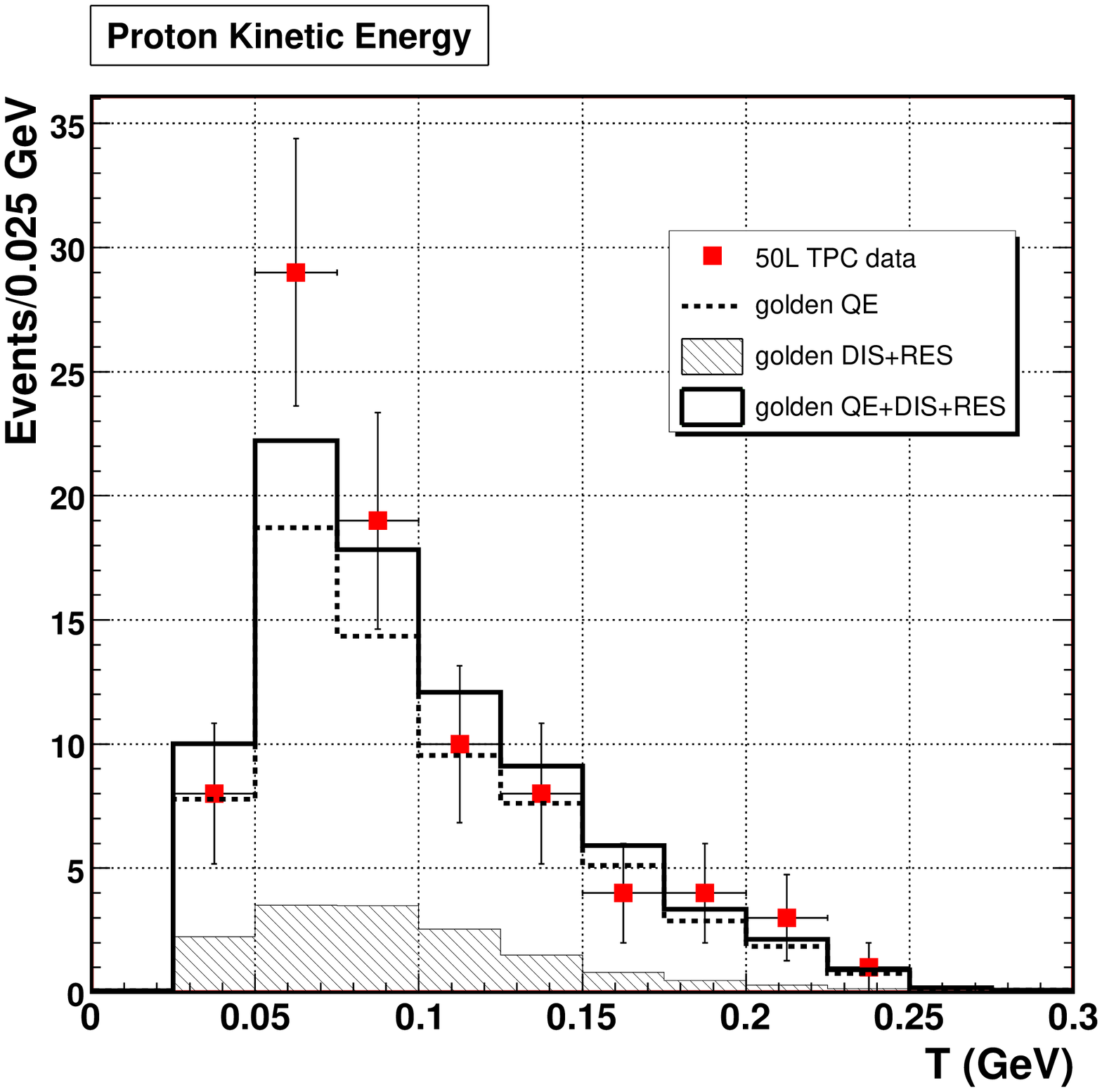} &
    \epsfysize=7.cm\epsfxsize=6.8cm\epsffile{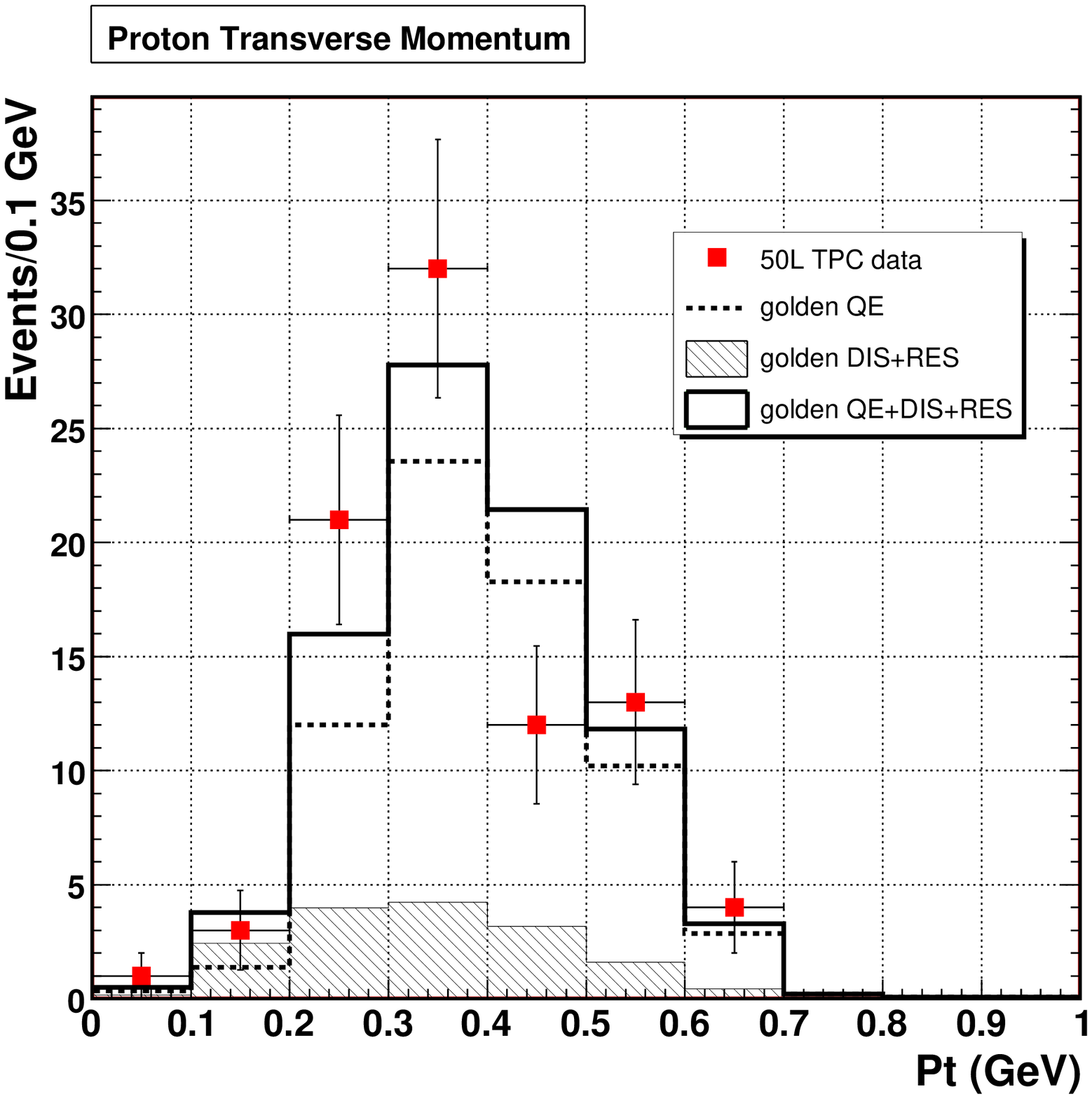} 
  \end{tabular}		
\caption{Distribution of the proton kinetic energy for the ``golden
sample'' (squared marks). 
Dotted histogram represents the expectation from simulated Monte-Carlo QE events, 
while the hatched one represents the simulated background from non-QE events. 
Both contributions are summed in the filled histogram and normalized to the golden sample data. }
\label{fig:proton_var}
\end{center}
\end{figure*}

\begin{figure*}[htbp]
\begin{center}
  \begin {tabular}{cc}
    \epsfysize=7.cm\epsfxsize=6.8cm\epsffile{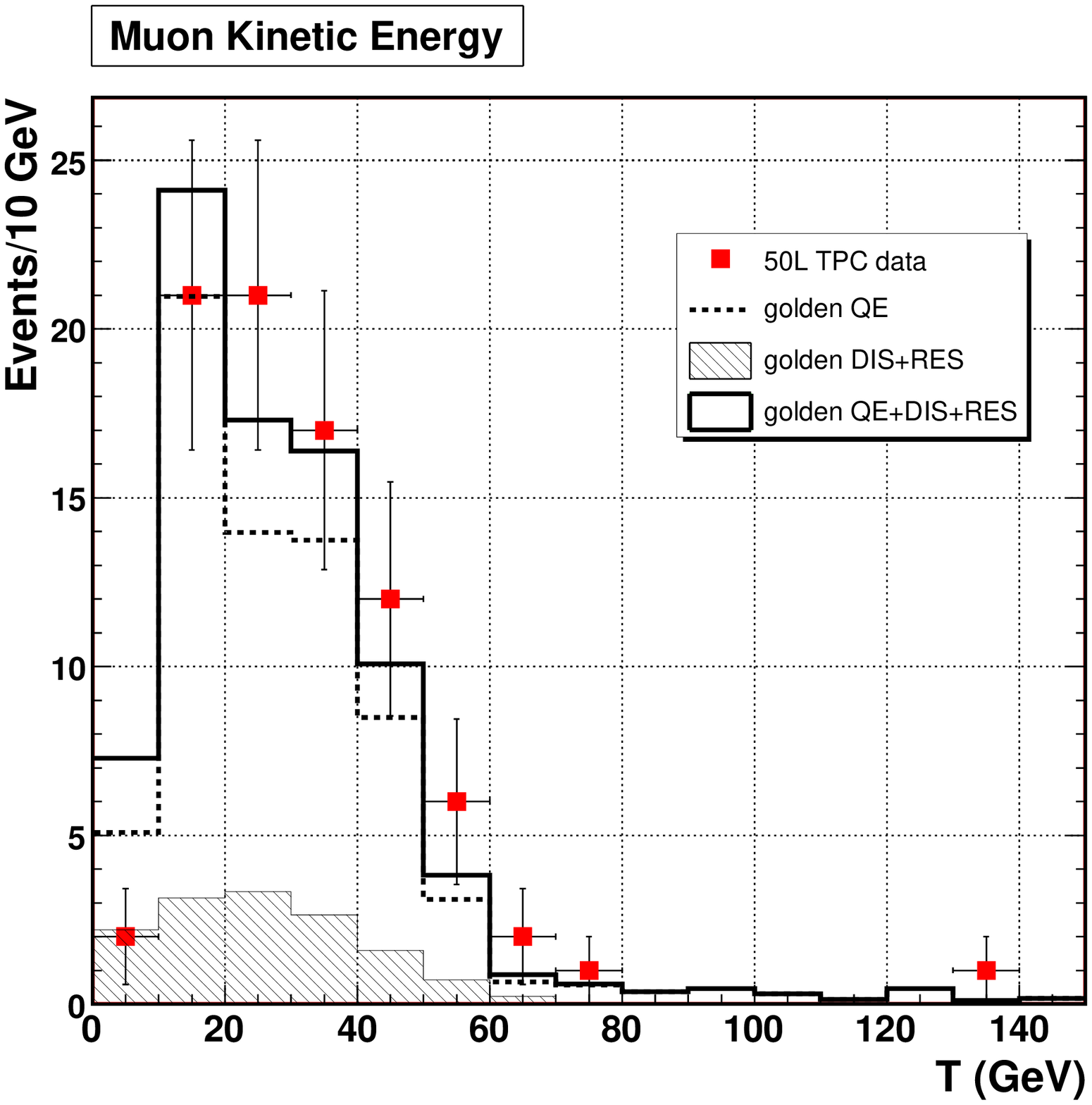} &
    \epsfysize=7.cm\epsfxsize=6.8cm\epsffile{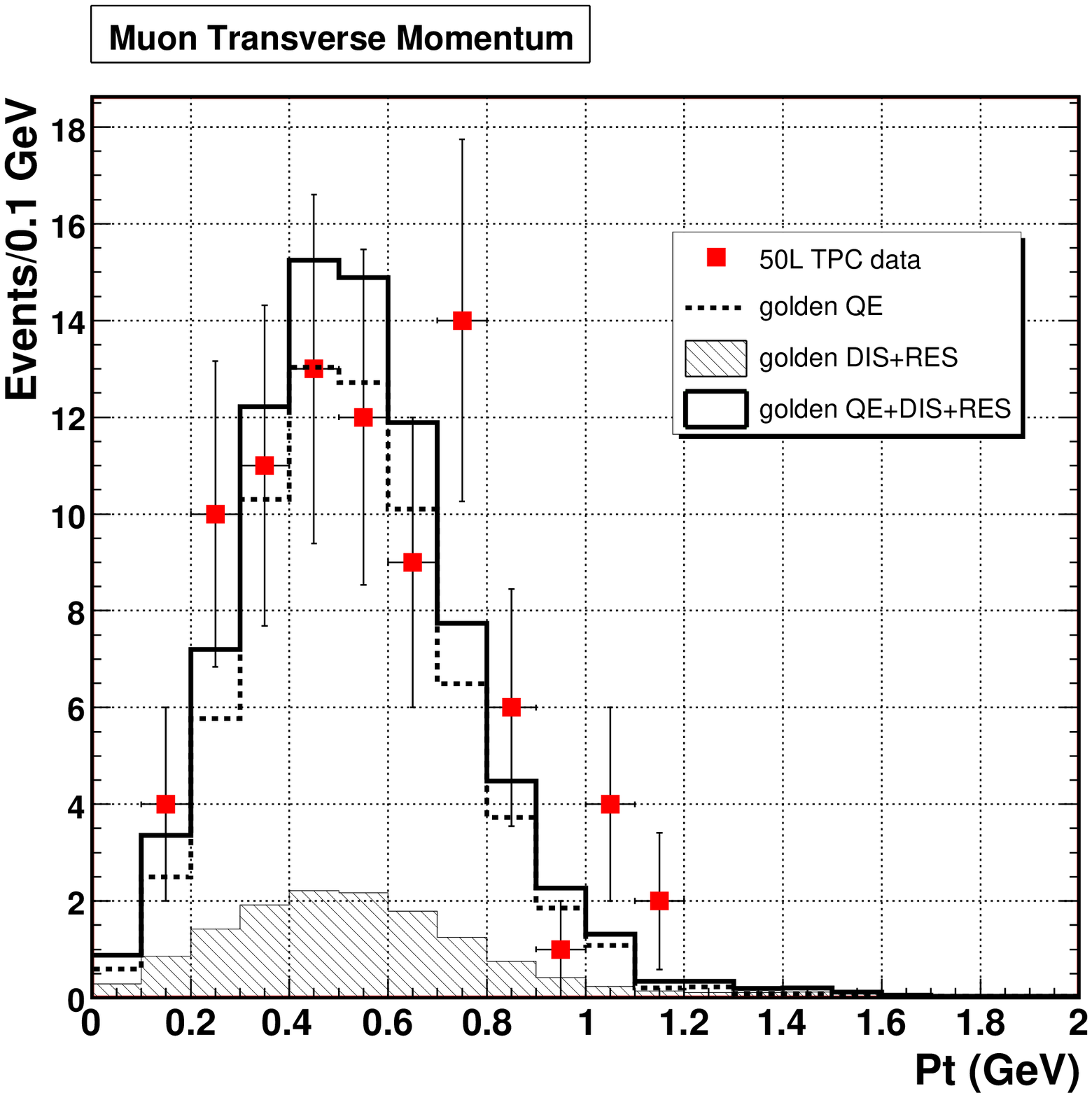} 
  \end{tabular}	
\caption{Distribution of the muon energy and the muon transverse momentum 
for the ``golden sample''. 
Dotted histogram represents the expectation from simulated Monte-Carlo QE events, 
while the hatched one represents the simulated background from non-QE events. 
Both contributions are summed in the filled histogram and normalized to the golden sample data. }
\label{fig:muon_var}
\end{center}
\end{figure*}

In spite of the limited statistics (86 events), the ``golden sample'' provides
information on the basic mechanisms that modify the kinematics of
neutrino-nucleus interactions with respect to the corresponding
neutrino-nucleon process. Nuclear matter perturbs the initial state of
the interaction through Fermi motion; it also affects the formation of
the asymptotic states through nuclear evaporation, hadronic
re-scattering or hadronic re-absorption. Several kinematic variables 
are only marginally affected by nuclear effects; in this case, the
corresponding distributions can be reproduced once the $\nu$-nucleon
interaction is corrected for Fermi motion and Pauli blocking.
Clearly, purely leptonic variables belong to this category. 
There are also a few hadronic variables whose distribution is strongly influenced by
the selection cuts but show limited sensitivity to nuclear effects. In
particular, the proton kinetic energy is bounded by the $T_p>40$~MeV
cut and the requirement of full containment in the fiducial volume.
This distribution is shown in Fig.~\ref{fig:proton_var} together with
the transverse momentum of the proton for the ``golden sample''.
Similarly, in Fig.~\ref{fig:muon_var} we show the kinetic energy
and the transverse momentum distributions of the muon. 
Figs.~\ref{fig:proton_var} and \ref{fig:muon_var} can
be used as a consistency check. They demonstrate that Monte-Carlo reproduces
the kinematic selection performed during the scanning and analysis of
the ``golden channel''. They also show that our selected sample contains 
a non-negligible contamination from non-QE events. As already indicated, we 
estimate this contamination to be 20~\%.

\begin{figure*}[htbp]
\begin{center}
  \begin {tabular}{cc}
    \epsfysize=7.cm\epsfxsize=6.8cm\epsffile{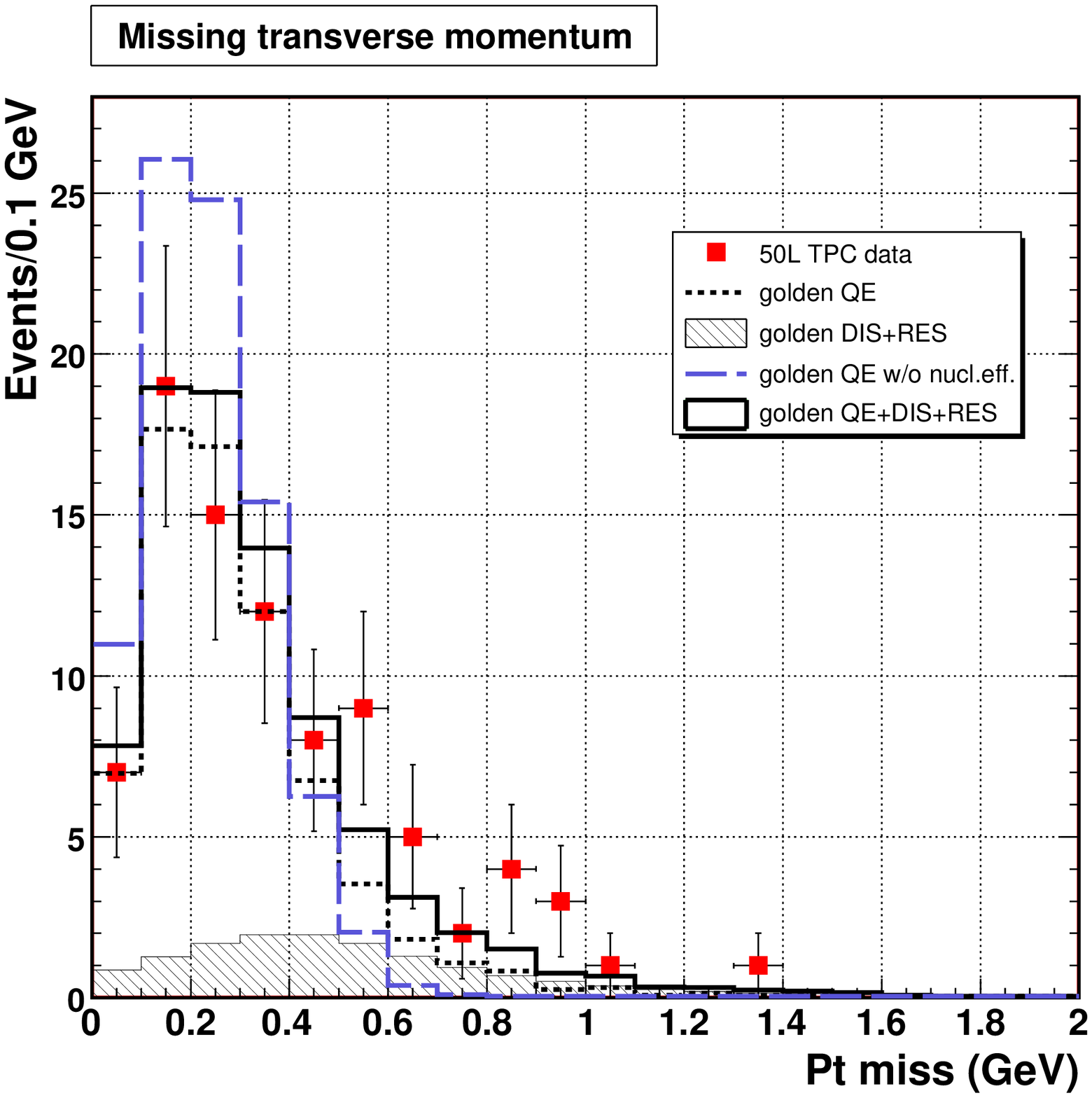} &
    \epsfysize=7.cm\epsfxsize=6.8cm\epsffile{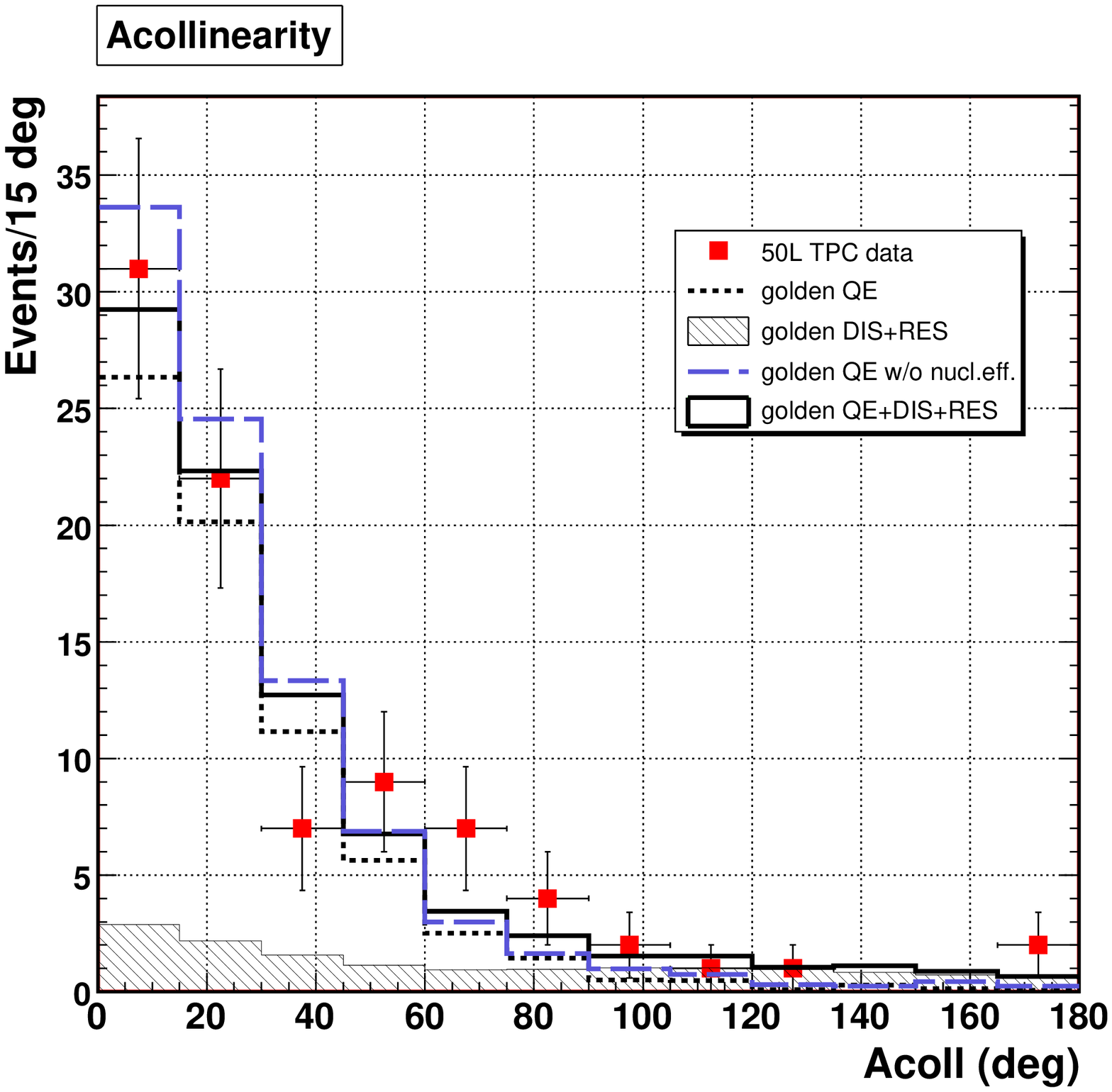} 
  \end{tabular}	
\caption{Distribution of the missing transverse momentum (left) and
acollinearity (right) for the ``golden sample''. 
Dotted histogram represents the expectation from simulated Monte-Carlo QE events, 
while the hatched one represents the simulated background from non-QE events. 
Both contributions are summed in the filled histogram and normalized to the golden sample data. 
The dashed histograms show the expected distributions in case no nuclear
effects are taken into account.}
\label{fig:event_var}
\end{center}
\end{figure*}

Other variables embedding the reconstructed kinematics of the protons
are sensitive to genuine nuclear effects. In particular, 
we analyzed two of them: the acollinearity and the missing transverse momentum of the event.
The former is defined as
\begin{equation}
A \equiv \mathrm{acos} \left[ \frac{ p_{xp} \ p_{x\mu} + p_{yp} \ p_{y\mu} }
{\sqrt { (p_{xp}^2 + p_{yp}^2) (p_{x\mu}^2 + p_{y\mu}^2) } } \right]
\end{equation}
$p_{xp}$ and $p_{yp}$ being the transverse momentum components of the
proton and $p_{x\mu}$ and $p_{y\mu}$ the corresponding quantities for
the muon. For a purely QE scattering on a nucleon, the muon and the
proton ought to be back-to back in the transverse plane so that the
acollinearity is zero. Fig.~\ref{fig:event_var} (right) shows the
acollinearity distribution for the ``golden sample'' and the
expectation from simulations. For this particular variable, the inclusion of nuclear effects
in the Monte-Carlo does not show a striking difference with respect to the case where
no nuclear effects are taken into account. The selection cuts (in particular, the fact that 
the proton should be fully contained), the resolution NOMAD has for muon reconstruction 
and the contamination from non-QE events are the main reasons for the appearance of events
with large acollinearity values. The distortion introduced in the tail of the acollinearity 
distribution by nuclear effects is much smaller than the one due to detector effects.
Hence this variable is not adequate to show how the event kinematics varies in the presence 
of nuclear matter.

The influence of nuclear effects on the event kinematics
is best seen when we consider the missing transverse momentum $p_T^{\mathrm{miss}}$ 
(see left plot on Fig.~\ref{fig:event_var}).
We observe that a naive approach that takes into account Fermi motion
and Pauli blocking disregarding nuclear effects (dashed histograms in 
Fig.~\ref{fig:event_var} left) does not reproduce the $p_T^{\mathrm{miss}}$ data.
Once nuclear effects are added, simulated events show a good agreement with data. 

In summary, we have seen that nuclear effects are an important 
source of perturbation for the kinematics of quasi-elastic neutrino interactions, 
however the smallness of the accumulated statistics does not allow a systematic survey of 
these effects~\cite{ref:nuint}. Still, these results empirically demonstrate the 
effectiveness of LAr imaging in the characterization of low-multiplicity multi-GeV 
neutrino interactions. In particular, a good description of transverse variables is
fundamental since they are key to a kinematic-based approach for 
$\nu_\mu~\rightarrow~\nu_\tau(\nu_e)$ oscillation appearance \cite{ref:oscsearch}.

\section{Conclusions}
We discussed the first exposure of a LAr TPC to a multi-GeV
neutrino beam. The data collected at the CERN WANF exploited the 3D
reconstruction capabilities of the LAr technology with
low-multiplicity events. The instrumental performances (size and stability of the
electron lifetime, reproducibility of the charge response for
mip and highly ionizing particles) and current off-line reconstruction tools
show that LAr TPC allows for an excellent particle identification and 
a precise measurement of the event kinematics.
In spite of the limited size of the detector, nuclear effects beyond Fermi motion
and Pauli blocking (e.g. hadron re-scattering or pion absorption) have
been observed as perturbations of the quasi-elastic $\nu_\mu$ CC interaction kinematics.
Variables in the transverse plane, which are the most sensitive to nuclear effects 
and key for neutrino oscillation searches, are well reproduced.

\section{Acknowledgments}
We wish to express our gratitude to the members of the NOMAD and
CHORUS collaboration for their support during the setting-up of the
test and data taking. We specially thank the help of the NOMAD Collaboration 
which agreed to increase slightly its dead-time to include the new dedicated trigger 
needed to acquire the escaping muons. We acknowledge the CERN SPS accelerator and
beam-line staff for the excellent performance of the WANF neutrino
beam. We thank the CERN PH Division since a large fraction of the LAr consumption 
costs were financially supported by it. The spanish group has been supported 
by M.E.C. through grant FPA2002-01835. The Polish authors acknowledge the support of 
the State Committee for Scientific Research in Poland, 105, 160, 620, 621/E-344, 
E-340, E-77, E-78/SPS/ICARUS/P-03/DZ211-214/2003-2005.

\end{document}